\documentclass[useAMS,usenatbib]{mn2e}
\usepackage[a4paper,colorlinks=True,citecolor=blue]{hyperref}
\usepackage{amssymb}
\usepackage{graphicx}
\usepackage[usenames,dvipsnames]{xcolor}
\usepackage{epstopdf}
\newcommand{\bea}{\begin{eqnarray}}
\newcommand{\eea}{\end{eqnarray}}
\newcommand{\beq}{\begin{equation}}
\newcommand{\eeq}{\end{equation}}

\newcommand{\sm}{\small}

\title[Testing cosmology with BOSS voids]{Testing cosmology with a catalogue of voids in the BOSS galaxy surveys}
\author[S. Nadathur]{Seshadri Nadathur$^{1}$\thanks{seshadri.nadathur@port.ac.uk}\\
$^1$Institute of Cosmology and Gravitation, University of Portsmouth, Burnaby Road, Portsmouth PO1 3FX, UK\\
}

\begin{document}

\date{\today}

\pagerange{\pageref{firstpage}--\pageref{lastpage}}

\label{firstpage}

\maketitle

\begin{abstract}
We present a public catalogue of voids in the Baryon Oscillation Spectroscopic Survey (BOSS) Data Release 11 LOWZ and CMASS galaxy surveys. This catalogue contains information on the location, sizes, densities, shapes and bounding surfaces of 8956 independent, disjoint voids, making it the largest public void catalogue to date. Voids are identified using a version of the {\sm ZOBOV} algorithm, the operation of which has been calibrated through tests on mock galaxy populations in $N$-body simulations, as well as on a suite of 4096 mock catalogues which fully reproduce the galaxy clustering, survey masks and selection functions. Based on this, we estimate a false positive detection rate of 3\%. Comparison with mock catalogues limits deviations of the void size distribution from that predicted in the $\Lambda$CDM model to be less than 6\% for voids with effective radius $8<R_v<60\,h^{-1}$Mpc and in the redshift range $0.15<z<0.7$. This could tightly constrain modified gravity scenarios and models with a varying equation of state, but we identify systematic biases which must be accounted for to reduce the theoretical uncertainty in the predictions for these models to the current level of precision attained from the data. We also examine the distribution of void densities and identify a deficit of the deepest voids relative to $\Lambda$CDM expectations, which is significant at more than the $3\sigma$ equivalent level. We discuss possible explanations for this discrepancy but at present its cause remains unknown.
\end{abstract}

\maketitle

\begin{keywords}
cosmology: observations -- large-scale structure of Universe -- methods: numerical -- methods: data analysis
\end{keywords}

\section{Introduction}
\label{sec:Intro}

Cosmic voids are large underdensities in the matter distribution which can be identified in galaxy redshift catalogues as regions of space containing fewer galaxies than average. The low matter density within voids means that they provide environments which are particularly sensitive to the effects of dark energy \citep[e.g.,][]{Bos:2012,Pisani:2015}, to modifications of the theory of gravity \citep{Li:2012,Clampitt:2013,Zivick:2015,Cai:2015,Barreira:2015}, or to other alternative cosmological models \citep{Yang:2014,Massara:2015}. Voids have been used in studies of the integrated Sachs-Wolfe effect \citep{Granett:2008a,Cai:2013ik,Hotchkiss:2015a,Kovacs:2015,Granett:2015}, as well as of weak gravitational lensing \citep[e.g.,][]{Clampitt:2015}.

All such studies require large and reliable catalogues of voids in galaxy surveys. A number of void catalogues have been compiled with data from the Sloan Digital Sky Survey (SDSS) Data Release 7 (DR7) \citep{Pan:2012,Sutter:2012wh,Ceccarelli:2013,Nadathur:2014a}, although different authors have used different methods and definitions of voids in each case. More recently the DR11 and DR12 releases from SDSS-III Baryon Oscillation Spectroscopic Survey (BOSS) have covered much larger volumes of the Universe, over a higher redshift range. \citet{Kitaura:2016} used a sample of voids drawn from BOSS DR11 data to measure the baryon acoustic oscillation (BAO) scale. Very recently, \citet{Mao:2016} have presented a catalogue of voids in the BOSS DR12 CMASS and LOWZ galaxy samples, which they used to compare the stellar mass distribution of galaxies inside and outside voids.

In this paper, we present a new public catalogue of voids drawn from the BOSS DR11 data, which is complementary to those described above, and can be used for a variety of void studies. We consider disjoint voids -- independent underdense regions of space that do not overlap with each other -- obtained from a modified version of the popular {\sm ZOBOV} \citep{Neyrinck:2008} void-finder, which does not impose any a priori conditions on the void shape. We provide a wealth of information on each void, including locations, sizes, densities, shape parameters and bounding surfaces. To calibrate the operation of this algorithm and to estimate the matter content of the voids thus obtained, we compare to the characteristics of voids found in mock galaxy populations in $N$-body simulations. To beat down cosmic variance and to account for systematic effects in the data, we also compare to a large suite of 4096 mock galaxy catalogues obtained from augmented Lagrangian perturbation theory and matching all characteristics of the data DR11 samples including the survey masks and selection \citep{Kitaura:2013,Kitaura-DR12mocks:2016}.

The catalogue presented here contains a total of 8956 independent voids, with an estimated false positive detection rate of 3\%. This makes it the largest public catalogue of voids to date, and provides much greater statistical power for void studies. The difference in catalogue size compared to that of \citet{Mao:2016}, to whom our void-finding approach is broadly similar, is that they use a much more restrictive, and in our opinion unnecessary, final selection cut. As a result the size of our catalogue exceeds theirs by more than a factor of 7, with a corresponding increase in the statistical power of constraints obtained from it.

We utilise this statistical power to compare the properties of voids in our catalogue to theoretical expectations from the standard $\Lambda$ Cold Dark Matter ($\Lambda$CDM) model. Of particular interest is the distribution of void sizes, which has been proposed as sensitive probe of a range of alternative scenarios, including modified gravity models \citep[e.g.,][]{Zivick:2015,Cai:2015}, and models with a varying equation of state for dark energy \citep[e.g.,][]{Pisani:2015}. Using data from our void catalogue, we are able to constrain deviations of this key observable from the $\Lambda$CDM value to be $<6\%$ (at 95\% confidence) over for void sizes up to an effective radius of $60\,h^{-1}$Mpc, with the constraint from the higher-redshift CMASS sample alone being even tighter. The void data are in excellent agreement with theoretical expectation throughout. 

These constraints are much tighter than the deviations predicted by the alternative models considered in the works above. However, we also note important systematic effects on the void size distribution due to galaxy bias and survey boundary effects, which have not been accounted for in these models. These theoretical uncertainties will need to be reduced to match the precision that can be obtained from the data presented in this work.

We also compare the distribution of void densities to the $\Lambda$CDM prediction. This is a complementary statistical test to that provided by the size distribution, and in fact we find a deficit of voids with the deepest density minima relative to expectations. This deficit ranges from $\sim20\%$ to $\sim50\%$ over a range of void densities, and is significant at more than the $3\sigma$ equivalent confidence level. We discuss various possible explanations for this discrepancy, including a previously undetected systematic effect in the mock catalogues, and physical models which could might produce the qualitatively correct effect of shallower voids.

The layout of the paper is as follows. In Sec.~\ref{sec:data&mocks} we present details of the BOSS data samples and the mock galaxy catalogues used in this analysis. Sec.~\ref{sec:methods} presents our methods: the void-finding algorithm and treatment of survey data are presented in Secs.~\ref{subsec:ZOBOV} and \ref{subsec:mask}, the calculation of void properties in Sec.~\ref{subsec:voidproperties}, and the calibration against mock catalogues in Sec.~\ref{subsec:BigMDmocks}. Results from the void catalogue are presented in Sec.~\ref{sec:results}, and the cosmological tests in Sec.~\ref{sec:deviations}. Finally we provide conclusions and a future outlook in Sec.~\ref{sec:conclusions}. Our void catalogue and associated materials are made available for public download.\footnote{\url{http://www.icg.port.ac.uk/stable/nadathur/voids/}}

\section{Data samples and mocks}
\label{sec:data&mocks}

\subsection{BOSS survey data}
\label{subsec:BOSSdata}

The galaxy data used in this work is taken from the SDSS-III BOSS Data Release 11 (DR11) \citep{Alam-DR11&12:2015}. BOSS has obtained spectra for over 1.37 million galaxies in two  contiguous regions of the sky in the Northern and Southern Galactic Caps (referred to hereafter as North and South), which in total cover slightly more than one-fifth of the sky (over $10,000\;\rmn{deg}^2$). Target selection for the LOWZ ($z<0.45$) and CMASS ($0.4<z<0.7$) samples is aimed at producing two galaxy catalogues for large-scale structure studies covering different redshift ranges. Details of the target selection, data reduction algorithms and catalogue creation are presented in \citep{Eisenstein-BOSS:2011,Dawson:2013,Alam-DR11&12:2015,Reid-DR12:2016}. The latest DR12 release increases the sky coverage by approximately $10\%$ over the publicly available DR11 data considered in this work.\footnote{\url{http://data.sdss3.org/sas/dr11/boss/lss/}}

\begin{figure*}
\begin{center}
\includegraphics[width=82mm]{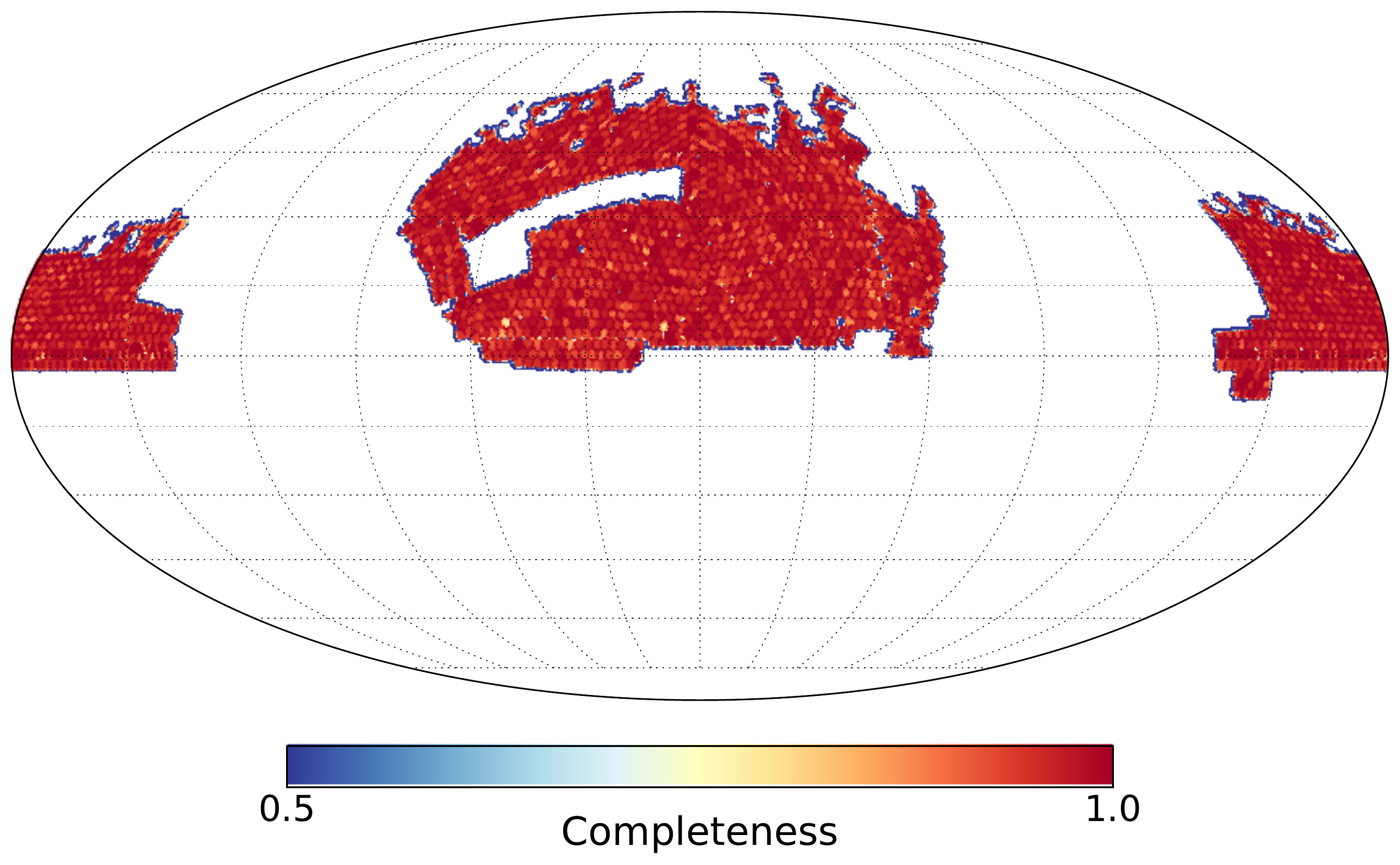}
\includegraphics[width=82mm]{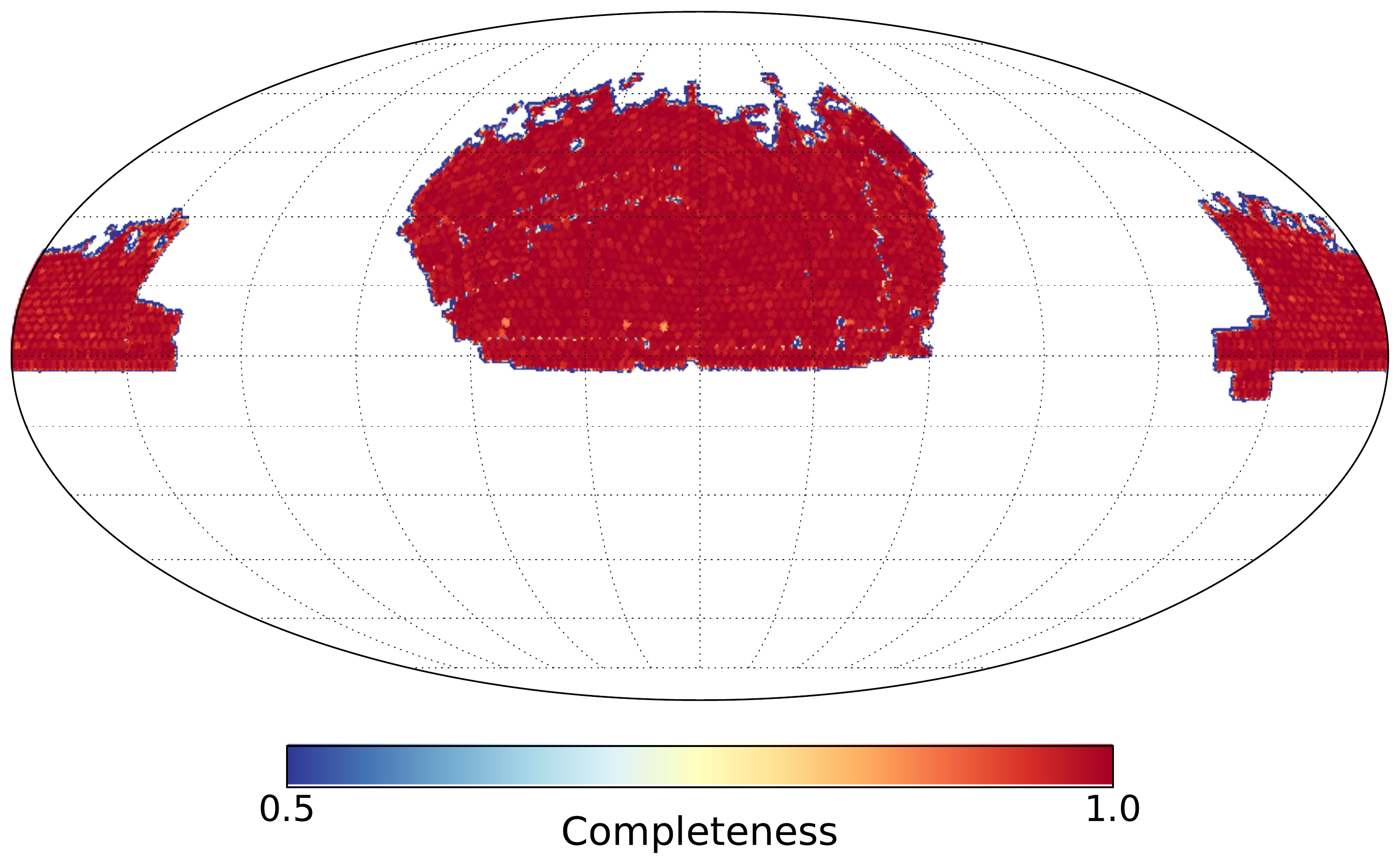}
\caption{Survey masks and angular completeness for the LOWZ (left) and CMASS (right) surveys, showing both Northern and Southern Galactic regions. These masks are accounted for by placing of buffer mocks with 10 times the galaxy number density of the survey around the survey edges and in interior holes to limit the leakage of Voronoi cells outside the surveyed volume. The density estimate derived from the tessellation is limited to regions interior to the mask, as described in the text.}
\label{fig:surveymasks}
\end{center}
\end{figure*}

To the LSS catalogues provided by BOSS, we impose the following redshift cuts: $0.15<z<0.43$ (LOWZ) and $0.43<z<0.7$ (CMASS). The survey masks and angular completeness of the resultant samples are shown in Figure~\ref{fig:surveymasks}. %and the redshift-dependent selection functions in Figure~\ref{fig:selfns}. 
The selection function for both surveys is also redshift-dependent, leading to significant variation in the local mean galaxy number density $n(z)$ over the survey redshift ranges. This is accounted in the void-finding for as described in Sec.~\ref{subsec:mask}. For each of the LOWZ and CMASS catalogues, the North and South regions are treated separately for void-finding purposes, but the catalogues are combined for the statistical results presented below. In all, the LOWZ survey covers a volume of $1.15\,h^{-3}$Gpc$^3$, while CMASS covers $3.41\,h^{-3}$Gpc$^3$.

%==================Fig.: =======================%
%==================Fig.:=======================%

%%==================Fig.: =======================%
%\begin{figure*}
%\begin{center}
%\includegraphics[width=70mm]{../../figures/DR11/LOWZ_selfns.pdf}
%\includegraphics[width=70mm]{../../figures/DR11/CMASS_selfns.pdf}
%\caption{Radial selection functions for the LOWZ (left) and CMASS (right) galaxy data, as a function of survey redshift $z$. North and South galactic regions are shown separately. This variation of the mean number density of observed galaxies is accounted for in determining the local density from the tessellation information, as described in Sec.~\ref{subsec:mask}.}
%\label{fig:selfns}
%\end{center}
%\end{figure*}
%%==================Fig.:=======================%

\subsection{Mocks}
\label{subsec:mocks}

Although a theoretical model of voids exists \citep{Sheth:2003py} and has attractive features, including analytic tractability, it unfortunately fails -- by orders of magnitude -- to correctly describe the statistics of voids actually found in data \citep{Nadathur:2015b}. In addition, the role of galaxy bias in affecting void properties has been highlighted by \citet{Nadathur:2015c}. Therefore, in order to understand the action of the void-finding algorithm on galaxy survey data and to calibrate expectations, one must make use of mock galaxy catalogues in simulations.

We make use of two types of mock catalogues in this work. The first is a set of lightcone mock galaxy catalogues for the DR11 data samples \citep{Kitaura-DR12mocks:2016}, constructed using the {\sm PATCHY} code \citep{Kitaura-PATCHY:2014} and publicly available from SDSS.\footnote{\url{http://data.sdss3.org/sas/dr11/boss/lss/dr11_patchy_mocks/}}. These consist of 4096 mock catalogues, i.e., 1024 each for the CMASS North, CMASS South, LOWZ North and LOWZ South samples, and include all aspects of the survey angular completeness and selection function. The {\sm PATCHY} algorithm is based on matching the large-scale density field obtained from augmented Lagrangian perturbation theory \citep{Kitaura:2013} rather than a full $N$-body prescription, and then populating the simulation with mock galaxies using a halo abundance matching (HAM) technique. The pipeline produces mocks on a light-cone, and reproduces the actual survey geometry, sector completeness, veto masks and radial selection functions. \citet{Kitaura-DR12mocks:2016} show that the resulting mock catalogues accurately reproduce both the two-point and three-point statistics of the galaxy distribution, thus providing as close as possible a realisation of the observed data. We will refer to these as the {\small PATCHY} mocks hereafter.

In addition to these we also use halo catalogues from the Big MultiDark (BigMD) $N$-body simulation \citep{Klypin:2014}. This simulation consists of $3840^3$ particles in a box of side $2.5\;h^{-1}$Gpc, evolved using the {\small GADGET-2} \citep{Springel:2005} and Adaptive Refinement Tree (ART) \citep{Kravtsov:1997,Gottloeber:2008} codes, with cosmological parameters $\Omega_M=0.307$, $\Omega_B=0.048$, $\Omega_\Lambda=0.693$, $n_\rmn{s}=0.95$, $\sigma_8=0.825$ and $h=69.3$. The simulation initial conditions are set using the Zeldovich approximation at starting redshift $z_i=100$. The box volume of BigMD thus exceeds that of the LOWZ survey by a factor of $\sim14$, and that of CMASS by a factor of $\sim5$.

Haloes are found in this simulation using the Bound Density Maximum algorithm \citep[BDM][]{Klypin:1997,Riebe:2013}. To model the LOWZ and CMASS galaxy samples, we make use of the halo catalogues from two redshift slices, at $z=0.32$ and $z=0.52$ respectively. These haloes are then populated with galaxies according to the Halo Occupation Distribution (HOD) model of \citet{Zheng:2007}, using parameters determined by \citet{Manera:2015} and \citet{Manera:2013} for the LOWZ and CMASS galaxy samples, in order to approximately match the mean galaxy number density and galaxy bias for the DR11 data. The mock catalogues thus constructed are henceforth referred to as the BigMD LOWZ and CMASS mocks.

Thus the BigMD HOD mock samples differ from the {\sm PATCHY} mocks primarily because they fill a cubic simulation box with periodic boundary conditions rather than mimicking the survey boundary. This is an advantage, because it allows us to directly isolate the effects of the finite survey extent and the complex mask with holes (Figure~\ref{fig:surveymasks}) on void observables and statistics. Most previous studies the cosmological constraints achievable from void statistics in future surveys \citep[e.g.,][]{Pisani:2015,Zivick:2015,Cai:2015} have ignored the role of the survey geometry. By a comparison of the BigMD and {\sm PATCHY} mocks we will show in Sec.~\ref{sec:results} that for some observables this introduces a large systematic bias in theoretical expectations which is already important for current data.

Finally, another advantage of using the BigMD simulation is that it allows access to the true underlying dark matter density fields and potentials in void regions, which we will make use of in calibrating our expectations for the matter content of galaxy voids. In particular, this is important for estimating the void `significance' and for optimising quality cuts to the final void catalogue, as discussed in Sec.~\ref{subsec:BigMDmocks}. To measure the dark matter density within galaxy voids in BigMD, we make use of a cloud-in-cell (CIC) density estimator on a $2350^3$ grid, determined from the full resolution simulation snapshot at each redshift. The $\gtrsim1\;h^{-1}$Mpc resolution of this grid is more than sufficient for typical void sizes in the mock catalogues.

\section{Voidfinder and methodology}
\label{sec:methods}

\subsection{Modified {\sm ZOBOV} void-finder}
\label{subsec:ZOBOV}

In this work we make use of a modified version of the {\small ZOBOV} void-finding algorithm \citep{Neyrinck:2008}. {\sm ZOBOV} works on the input discrete point set of galaxy positions by performing a Voronoi tessellation in order to partition the survey or simulation volume into Voronoi cells associated with each galaxy, each of which contains the region of space closer to that galaxy than to any other. The local density field within each cell is then estimated based on the inverse of the cell volume; this is known as the Voronoi Tessellation Field Estimator (VTFE) technique. {\sm ZOBOV} then operates on the reconstructed density field, identifying local minima and the watershed basins around them to produce a list of voids. In identifying these voids, the algorithm makes no prior assumptions about the void shape, instead respecting the true topology of underdensities in the galaxy distribution. 

An additional final step which is often performed is to merge neighbouring underdensities together according to the watershed principle to form a larger single void if they satisfy certain criteria. The problem with this procedure is that there is no unambiguous and widely agreed set of criteria to control this merging, and a wide variety of different criteria have been used in the past \citep[see, e.g.,][]{Neyrinck:2008,Sutter:2012wh,Nadathur:2014a,Nadathur:2015a,Cautun:2016,Mao:2016}. A comparison of the relative merits of certain choices and the effects they have on void parameters is provided by \citet{Nadathur:2015c}. Generally, the criteria chosen are strict enough that merging is rare, so the majority of voids remain unmerged. However, the properties of the very largest and deepest voids, which are often the ones of greatest interest and also the most likely to undergo merging, are very sensitive to the details of the merging criteria chosen. 

We therefore choose not to merge any neighbouring voids. By doing so, we retain as much information as possible about the underlying topology of the reconstructed density field. This choice also highlights a series of useful degeneracies between several void observables (see Sec.~\ref{subsubsec:degeneracies}). In preparing the void catalogue, we experimented with alternative choices of merging neighbouring voids (see also \citealt{Nadathur:2015c}), but these were found to introduce additional noise in these relationships, and in some cases to invert them, so were not preferred over the simplest choice implemented here.

\subsection{Accounting for the survey mask and selection function}
\label{subsec:mask}

\begin{figure*}
\begin{center}
\includegraphics[width=150mm]{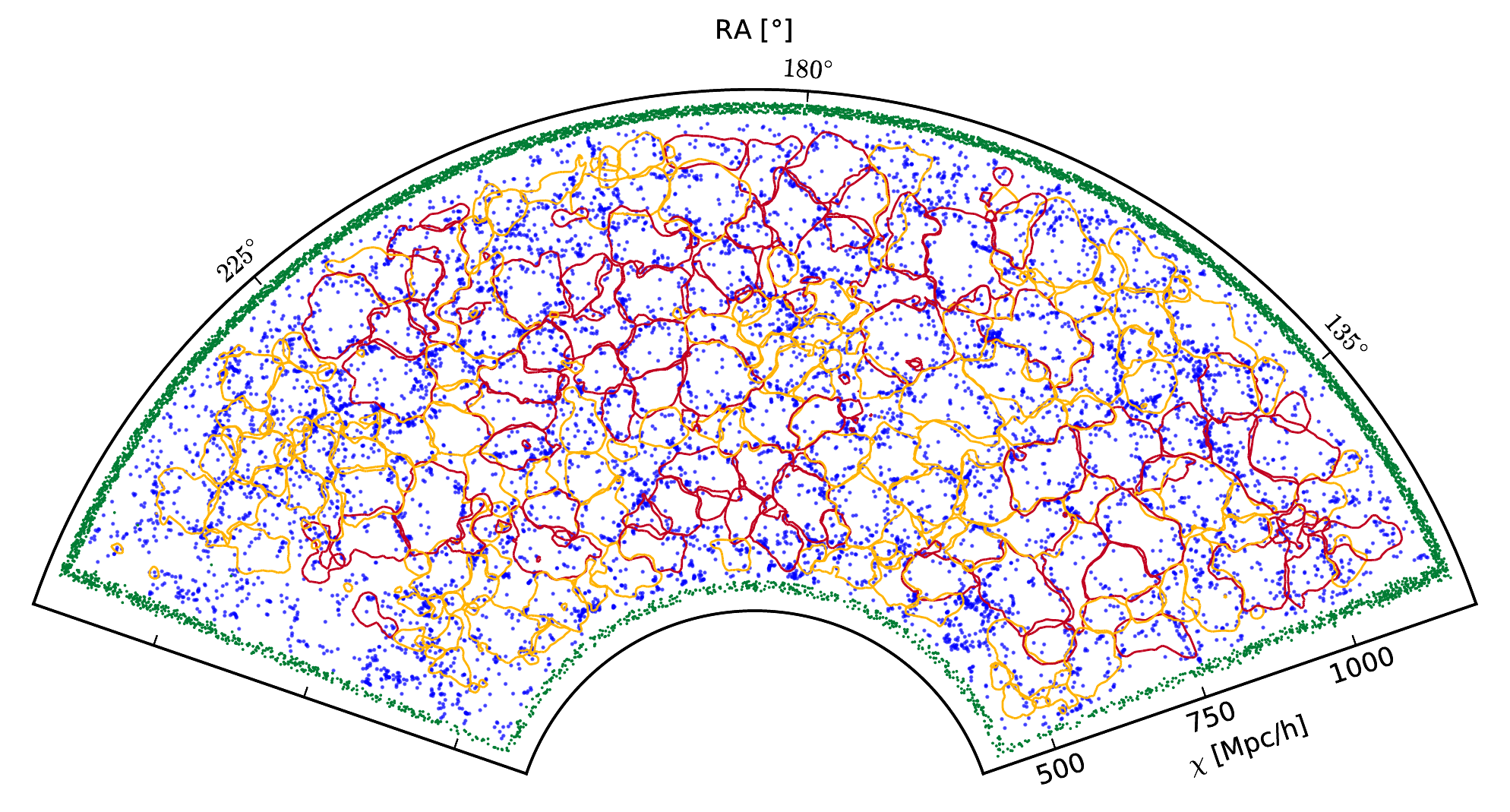}
\caption{The cross-section of voids in the LOWZ survey with the cone $\rmn{Dec}=12^\circ$, coloured according to whether the average galaxy density within the void is $\overline\delta_g<0$ (red) or $\overline\delta_g>0$ (yellow). Galaxy positions in a slice of opening angle $2^\circ$ centred at the angle are overlaid in blue, and buffer mocks around the survey edges in green. Voids with $\overline\delta_g<0$ tend to correspond to under-compensated underdensities, while those with $\overline\delta_g>0$ are on average over-compensated on large scales \citep{Nadathur:2015c}.}
\label{fig:decslice}
\end{center}
\end{figure*}

The {\sm ZOBOV} algorithm described above can be straightforwardly applied to the BigMD LOWZ and CMASS mocks due to the cubic geometry and periodic boundary conditions. In order to apply it to the DR11 galaxy data and the {\sm PATCHY} mocks, we broadly follow the algorithm described in detail by \citet{Nadathur:2014a}. We first convert (mock) galaxy positions from sky coordinates and redshift to Cartesian positions, using $\Omega_M=0.308$ based on the latest cosmological results from \citet{Planck:2015params}. To prevent the tessellation from leaking beyond the observed survey volume, we then place a thin layer of buffer particles around the boundary of the survey mask and along both the high- and low-redshift caps. Buffer particles are also placed inside large `holes' in the survey mask due to bright stars or other effects. To determine the survey boundary and holes, we use a rasterized HEALPix \citep{Gorski:2004by} version of the survey mask at resolution $N_\mathrm{side}=128$. The density of these buffer particles is 10 times that of the survey galaxies, and their placement is illustrated in Figure~\ref{fig:decslice}, which shows a thin slice through the LOWZ survey volume. The tessellation is then performed over a large cubic volume completely enclosing the survey and the buffer. In order to stabilise the tessellation algorithm during operation, sparse 'guard' particles are added to empty regions of the cube far from the survey volume.

%==================Fig.: =======================%
%==================Fig.:=======================%

%==================Fig.: =======================%
\begin{figure*}
\begin{center}
\includegraphics[width=75mm]{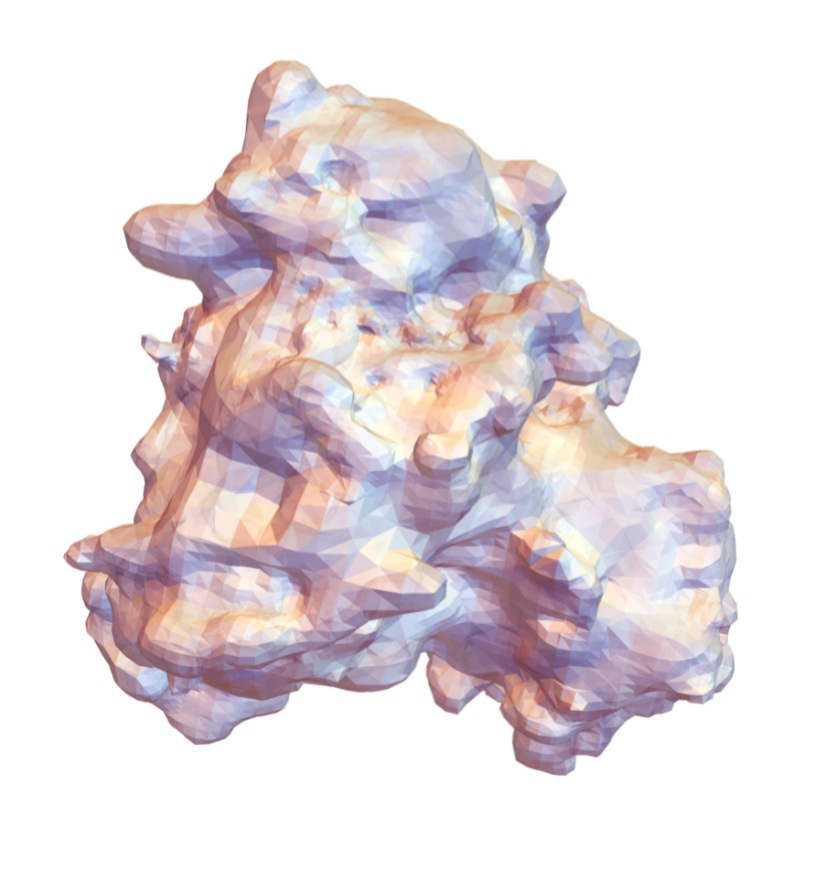}\hspace{2em}
\includegraphics[width=75mm]{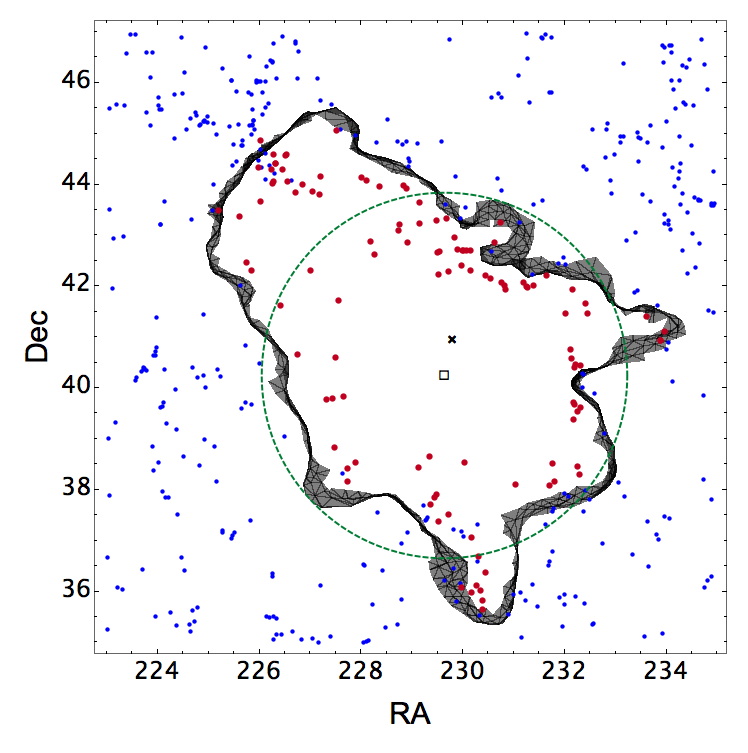}
\caption{\emph{Left}: A 3-dimensional representation of the bounding surface of an example void from the CMASS catalogue, obtained using the surface information described in Sec.~\ref{subsubsec:surfaces}. This void has effective radius $R_v=86.7\,h^{-1}$Mpc, effective ellipticity $e=0.05$, $\delta_{g,\rmn{min}}=-0.93$ and $\overline\delta_g=-0.18$. \emph{Right}: A thin slice through the void at redshift $z=0.528$ is shown by the black `ribbon'. Points show the projected positions of galaxies lying within a slice of thickness $30\,h^{-1}$Mpc centred at this redshift, with void member galaxies shown with the larger (red) points. The void minimum density centre and the volume-weighted barycentre of its member galaxies are shown by the black square and cross respectively. The green dashed line is the circle projected on the sky by an equivalent sphere of radius $R_v$.} 
\label{fig:examplevoid}
\end{center}
\end{figure*}
%==================Fig.:=======================%

After the tessellation has been carried out, the volume $V_i$ of the Voronoi cell of the $i$th galaxy is converted to a normalised density estimate via
\beq
\label{eq:Voronoi}
\frac{\rho_i}{\overline\rho} = \frac{\overline{V}}{V_i \phi(z_i)},
\eeq
where $\overline{V}$ is the mean volume for \emph{galaxy} cells (excluding those of the buffer and guard particles), and $\phi(z_i)$ represents the radial selection function at the redshift $z_i$.\footnote{In principle one could also weight the volumes to account for the varying angular selection function, but in practice given the small variation in completeness seen in Figure~\ref{fig:surveymasks} this was found to be unimportant.} {\sm ZOBOV} then uses this density estimate to find local minima and create the watershed voids. The small fraction of such local minima with minimum densities larger than the mean are rejected.

The addition of buffer mocks around the survey edge is necessary to prevent the VTFE estimator from assigning artificially low densities to galaxies near an edge. However, it also naturally distorts the density estimates of such edge galaxies. Any galaxy whose Voronoi cell is adjacent to that of a buffer particle is therefore identified and removed from the list before proceeding to the next step of identifying density minima and watershed zones; edge galaxy Voronoi volumes are never added to voids. This leads to the definition of two classes of voids. Any void which contains a galaxy that is itself adjacent to such an edge galaxy is flagged as an \emph{edge void}. The true extent of such voids is likely to have been artificially truncated by the survey edges. All other voids contain their complete volumes. For some cosmological purposes it may be necessary to restrict the sample to non-edge voids only; however we shall not do so in this work.

For a given survey geometry, the higher the density of particles in the buffer layer the larger the fraction of edge galaxies which must be discarded. However, it is still possible for a galaxy Voronoi cell to extend far outside the survey volume \emph{without} it being flagged as an edge galaxy, and this probability increases as the buffer density decreases. Such tessellation failures have led to instances of `voids' in previous catalogues that lie outside the survey mask.\footnote{For instance, see the discussion in footnote 6 of \citet{Kovacs:2015}.} Based on the {\sm PATCHY} mock realizations, we estimate that with the buffer density set to 10 times the true galaxy density, such a tessellation failure occurs for one of the $\sim520,000$ galaxies in CMASS in about 5\% of the mocks. We therefore include an \emph{extra} check after watershed algorithm stage to remove such spurious voids. 

When {\sm ZOBOV} is applied to a cubic simulation box, the resulting void catalogue is \emph{space-filling}: to a good approximation every volume element of the box is assigned to some void. However, this is not the case for survey data, primarily because of the fraction of unusable survey volume lost to edge galaxies. In addition, the complex survey geometry and holes can disrupt tessellation connectivity such that some density minima lie above the mean $\overline\rho$ and are rejected.

\subsection{Defining void properties}
\label{subsec:voidproperties}

Having obtained the catalogue of voids according the algorithm described above, we then obtain the following key observable properties for each void.

\subsubsection{Centre locations}
\label{subsubsec:centres}

We define the void centre to be the circumcentre of the positions of the lowest-density galaxy in the void and its three lowest-density mutually adjacent neighbours \citep{Nadathur:2015b}. This is equivalent to defining the centre of the largest empty sphere that can be inscribed in the void, and is thus the location of minimum galaxy density. In this sense it is very similar to the centre definition employed by the {\sm DIVE} void-finding algorithm \citep{Zhao:2016}, which also uses a tessellation-based density estimator. The difference between our implementation and that of \citet{Zhao:2016} is that we only report a single void and centre for each watershed region.

Another definition of the void centre commonly used in the literature \citep[e.g.,][]{Sutter:2012wh,Nadathur:2014a,Mao:2016} is the volume-weighted barycentre of the galaxies within the void, $\mathbf{X}_\rmn{bary}=\sum_i \mathbf{x}_iV_i/\sum_i V_i$. A visual example of the practical difference in the two definitions is shown in Figure~\ref{fig:examplevoid}. A statistical comparison of the two definitions was provided by \citet{Nadathur:2015b}, who showed that the minimum density centre is significantly better correlated with the true location of the matter underdensity within the void. Unless otherwise specified, all results in this paper will therefore refer to the void minimum density centre location, but barycentres are also provided in the accompanying public void catalogue.

\subsubsection{Sizes}
\label{subsubsec:sizes}

The total volume of each void is simply the sum of the volumes of its constituent Voronoi cells, $V_v = \sum_iV_i$. We define an effective void radius $R_v$, to be the radius of an equivalent sphere of this volume,
\beq
\label{eq:Rv}
R_v = \left(\frac{3}{4\pi}V_v\right)^{1/3}.
\eeq 
In general however individual void shapes are far from spherical, as can be seen from Figures~\ref{fig:decslice} and \ref{fig:examplevoid}.

\subsubsection{Densities}
\label{subsubsec:densities}

In addition to their sizes, voids can importantly be characterized by their densities, since they trace minima of different depths. We use five different measures of the void density, two of which can be directly determined from the galaxy distribution and three which refer to the total matter content of voids and therefore must be inferred from observables based on calibration with simulation results.

The minimum galaxy density within a void is represented by $\delta_{g,\rmn{min}}=\rho_{g,\rmn{min}}/\overline\rho-1$. This is simply the minimum density contrast estimated from the VTFE reconstruction, and is the value reported by {\sm ZOBOV} as the `core' density for each void.

The \emph{average} galaxy density contrast is defined as
\beq
\label{eq:deltag}
\overline\delta_g = \frac{1}{\mathcal{V}}\int_\mathcal{V}\frac{\rho(\mathbf{x})}{\overline\rho}\,\rmn{d}^3\mathbf{x} -1,
\eeq
where the integral is performed over the three-dimensional void volume $\mathcal{V}$. In practice this is easily estimated from the VTFE density reconstruction as the volume-weighted average density of the void,
\beq
\label{eq:deltag2}
\overline\delta_g = \frac{1}{\overline\rho}\frac{\sum_i \rho_iV_i}{\sum_iV_i}-1,
\eeq
where the sum runs over all Voronoi cells that make up the void volume. Both observable densities $\delta_{g,\rmn{min}}$ and $\overline\delta_g$ make use of the reconstructed densities (equation~\ref{eq:Voronoi}) and therefore naturally account for the survey selection functions and boundaries.

While the galaxy density is instructive, we ideally wish to characterise voids by their true matter content. The matter density contrast at the void centre location is denoted by $\delta_\rmn{min}$. The integrated density contrast $\Delta$ is defined in terms of the density under a spherical top-hat filter centred at the void centre,
\beq
\label{eq:Delta}
\Delta(R) = \frac{3}{R^3}\int_0^R \delta(r)r^2\,\rmn{d}r,
\eeq
where $r$ is the distance from the void centre. From the BigMD simulation data, we calculate this value on two different scales, $\Delta(R_v)$ and $\Delta(3R_v)$, for each simulation void.

\subsubsection{Bounding surfaces}
\label{subsubsec:surfaces}

The bounding surface of a void is simply the union of external faces of the Voronoi cells making up its volume. Knowledge of the location of this boundary has been shown to be important by \citet{Cautun:2016}, who demonstrate that for irregularly shaped voids, the stacked void lensing signal can be enhanced by a factor of $\sim2$ if the stack is created as a function of the distance \emph{to} the nearest void boundary, rather than the distance \emph{from} the void centre.

To reconstruct a void boundary, we find all galaxies lying within the void that are adjacent to a galaxy lying outside it. The external face of the Voronoi cell is then simply the section of the plane that perpendicularly bisects the line joining these two galaxies. Each such plane is represented by its normal and a point on the plane, and this information is written to file for all external faces for the void. From this it is possible to reconstruct a three-dimensional image of any void as well as its sky cross-section, as shown in Figure~\ref{fig:examplevoid}. To specify the boundary for a typical medium-to-large void requires a few thousand such planes. Surface information for each void is provided for download in the public catalogue.

\subsubsection{Shapes and ellipticities}
\label{subsubsec:ellipticities}

Given the irregularity of void shapes visible in Figures~\ref{fig:decslice} and \ref{fig:examplevoid}, modelling voids as spheres of radius $R_v$ is clearly a gross simplification. A somewhat better approximation is to model them as tri-axial spheroids instead. 

To do this we first construct a cloud of points on the bounding surface for each void defined above, and calculate the corresponding inertia tensor,
\bea
\label{eq:inertiatensor}
I_{xx} &=& \sum_i(y_i^2+z_i^2),\\
I_{xy} &=& \sum_ix_iy_i\,
\eea
and so on, where $(x_i,y_i,z_i)$ are the coordinates of the $i$th point in the point cloud, relative to the void centre. Let the eigenvalues of this tensor be $\Lambda_1$, $\Lambda_2$ and $\Lambda_3$, in ascending order. Then we define \citep{BBKS}
\beq
\label{eq:ellipticity}
e \equiv \frac{1}{2}\frac{\Lambda_3-\Lambda_1}{\sum_i \Lambda_i},
\eeq
and
\beq
\label{eq:prolateness}
p \equiv \frac{1}{2}\frac{\Lambda_3-2\Lambda_2+\Lambda_1}{\sum_i \Lambda_i},
\eeq
to be the ellipticity and prolateness of the void, respectively. The ratio of the longest to shortest axis of the model spheroid is
\beq
\label{eq:q}
q = \sqrt{\frac{\Lambda_3}{\Lambda_1}}.
\eeq
Our definition of the inertia tensor $I_{ij}$ differs from that used by some previous authors \citep[e.g.,][]{Sutter:2014b,Sutter:2014c} because we use positions of points on the void bounding surface rather than galaxies within the void to define it. Since galaxies often lie well inside the void (Figure~\ref{fig:examplevoid}), this provides a better representation of the true shape of the void volume. Note also that our definition of the ellipticity $e$ obtained from this tensor is different.

\subsubsection{Density ratio}
\label{subsubsec:densratio}

The final observable characteristic of individual voids considered in this paper is the \emph{density ratio} $r$, defined as the ratio of the lowest value of the galaxy density along the edge of the void's watershed basin to the minimum galaxy density at the void centre. This quantity was introduced by \citet{Neyrinck:2008}, who advocated that it be used as a quality cut to assess the void probability, or `significance'. This is because the {\sm ZOBOV} algorithm reports all local density minima as voids, and therefore will report the existence of spurious voids even in a pure Poisson point set. Shot noise in the galaxy distribution may similarly result in regions being classified as `voids' despite not corresponding to real underdensities in the underlying matter distribution. Recently this has been used as a void quality cut by \citet{Mao:2016}, who require a threshold $r>1.57$ for voids in their catalogue.

Unfortunately, the void probability estimate reported by \citet{Neyrinck:2008} is based on the distribution of $r$ values for purely spurious voids in Poisson noise; therefore, it refers only to the probability $P(r|\rmn{Poisson})$. In designing a quality cut for creation of a catalogue of genuine voids, however, the quantity of interest is actually $P(\rmn{Poisson}|r)$, i.e., the probability that a candidate void with given value of $r$ does not correspond to a genuine matter underdensity. 

This issue was discussed in \citet{Nadathur:2015b}, where by analysing voids in $N$-body simulations we showed that $r$ does not provide a good discriminant for such cases. We provide further tests of this using the BigMD simulation mocks in Sec.~\ref{subsubsec:significance} and Figure~\ref{fig:significance} below, and reach the same conclusion. Furthermore, the results of \citet{Nadathur:2014a} for voids in SDSS DR7 surveys show that survey geometry effects introduce an additional large scatter in $r$ (see Fig. 5 of that paper), which mean it is less robustly determined than the other void observables.

We therefore do not impose any a priori cut on candidate voids based on this measure, although $r$ values for all voids are reported in the public version of the catalogue.

%==================Fig.: =======================%
\begin{figure*}
\begin{center}
\includegraphics[width=175mm]{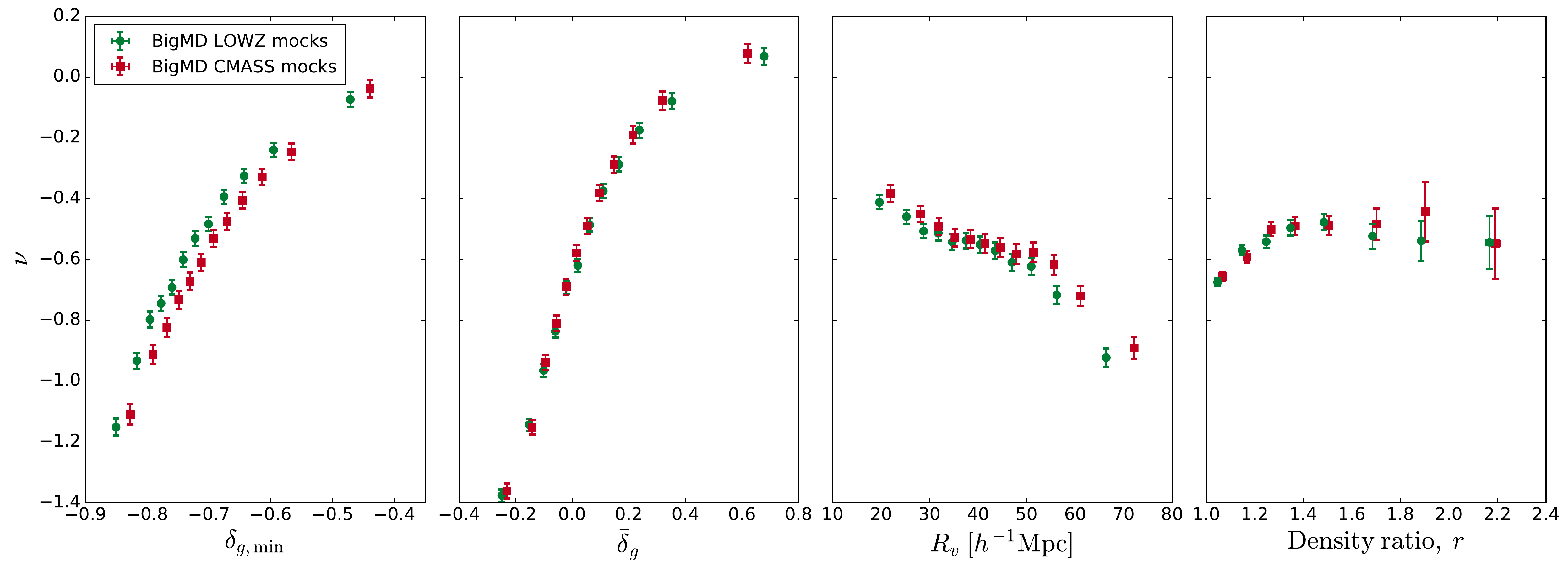}
\caption{The binned average void significance for voids in the BigMD simulation, measured in terms of the excursion $\nu=\Delta(R_v)/\sigma(R_v,z)$, as a function of different void observables: from left to right, minimum galaxy density $\delta_{g,\mathrm{min}}$, average galaxy density $\overline\delta_g$, size $R_v$, and density ratio $r$ (Sec.~\ref{subsec:voidproperties}). Smaller (more negative) values of $\nu$ are more significant, i.e., they correspond to rarer density fluctuations. $\overline\delta_g$ is the strongest predictor of void significance, $r$ is the weakest.}
\label{fig:significance}
\end{center}
\end{figure*}
%==================Fig.:=======================%

\subsection{Calibration with mocks}
\label{subsec:BigMDmocks}

The primary directly observable properties for each individual void -- its location, $R_v$, $\delta_{g,\rmn{min}}$, $\overline\delta_g$, $e$, $p$ (or $q$) and $r$ -- have been defined in the previous section. Before proceeding to the results from the DR11 data, we make use of voids in the BigMD mock catalogues to briefly review the relationships between these observables and the measures of the matter density within the void, $\delta_\rmn{min}$, $\Delta(R_v)$ and $\Delta(3R_v)$, that must be indirectly deduced from the galaxy distribution. From the BigMD mocks we obtain $4.3\times10^4$ LOWZ voids and $3.3\times10^4$ CMASS voids.

\subsubsection{Relationships between void observables}
\label{subsubsec:degeneracies}

The most striking relationship between void observables is that between $\delta_{g,\rmn{min}}$ and $R_v$: larger voids contain deeper density minima \citep{Nadathur:2015b,Nadathur:2015c}. This average relationship is universally true in simulations for all void sizes and for voids found using any tracer of the density field. It is also found to hold for voids found using other algorithms, such as the WVF \citep{Cautun:2016} and DIVE \citep{Zhao:2016} void-finders, so can be considered a generic property of voids. Interestingly, this behaviour is opposite to that predicted by the simplest versions of the excursion set void model of \citet{Sheth:2003py}.\footnote{The generic prediction within the \citet{Sheth:2003py} framework that smaller voids correspond to deeper density minima follows from the integral constraint assumption of that model and the statistics of an initially Gaussian random density field. For further details see \citet{Sheth:1999conf,Massara:2016}.}

Similarly, we find that the average ellipticity $e$ and prolateness $p$ of voids also decrease with $R_v$: larger voids are closer to sphericity. This is an intuitive result, which can be simply related to the theory of peaks of a Gaussian random field \citep{BBKS}. However, if some neighbouring watershed regions are allowed to merge to form a single larger void, the resultant is also less symmetrical about the centre and so the monotonicity in the $\langle e\rangle$-$\langle R_v\rangle$ relationship is lost for the very largest voids (here $\langle\cdot\rangle$ denotes the bin average over a subset of all voids). This adds to the complexity of modelling such structures and is a reason to disfavour void merging.

\subsubsection{Matter content of voids}
\label{subsubsec:voidmass}

\citet{Nadathur:2015b,Nadathur:2015c} examined the matter content of voids in simulation and found the following relationships to observables:
\begin{enumerate}
\item void minimum densities satisfy $\langle \delta_\rmn{min}\rangle \simeq A\langle \delta_{g,\rmn{min}}\rangle + C$ for constant $A$ and $C$, but do \emph{not} satisfy the average deterministic linear bias relation $\delta=\delta_g/b$,
\item similarly, $\langle \overline\delta_g\rangle\neq b\langle\Delta(R_v)\rangle$ and the average relationship in this case is not even linear, but
\item $\langle\overline\delta_g\rangle$ \emph{is} linearly proportional to $\langle\Delta(3R_v)\rangle$ for voids in any tracer population, with $\langle\overline\delta_g\rangle<0$ corresponding to $\langle\Delta(3R_v)\rangle<0$ and vice versa in every case.
\end{enumerate} 
Finally, these works also found $\langle\delta\rangle<0$ to hold for all subsets of the void population provided that the minimal quality cut $\delta_{g,\rmn{min}}<0$ was satisfied.

We confirm that all of these relationships hold true for voids in our BigMD LOWZ and CMASS mocks as well. We will provide a quantitative analysis and discussion of these relationships in forthcoming work \citep{Nadathur:2016c}, where we we will also show that the locations of voids with $\overline\delta_g<0$ are strongly correlated with peaks of the gravitational potential $\Phi$. This property might be particularly relevant for the use of voids for stacked ISW measurements \citep{Granett:2008a,Cai:2013ik,Hotchkiss:2015a,Kovacs:2015,Granett:2015}.

\subsubsection{Defining void significance}
\label{subsubsec:significance}

Based on this calibration of the void matter content, we wish to find a robust measure of the significance which may be attributed to each void. A simple way to do this is to describe the mass deficit within a void of size $R_v$ in units of the rms fluctuation of the matter density on that scale and at that redshift,
\beq
\label{eq:nu}
\nu = \frac{\Delta(R_v)}{\sigma(R_v,z)}.
\eeq
The value of $\nu$ then describes the depth of the excursion associated with the void. Note that $\nu$ could have been described on any scale. However, some choice is necessary and setting the scale to the void radius $R_v$ provides a link to an intuitive understanding of the mass deficit associated with the void.

In Figure~\ref{fig:significance}, we show the dependence of $\nu$ on different void observables, for voids in the BigMD LOWZ and CMASS mocks. The vast majority of voids correspond to $\nu<0$, i.e. genuine matter underdensities. The average galaxy density $\overline\delta_g$ is the strongest predictor of the void significance, closely followed by $\delta_{g,\rmn{min}}$. The weakest correlation is seen for the density ratio, $r$, indicating that using this quantity as a quality cut on void catalogues is suboptimal. 

It is worth emphasizing that the significance $\nu$ is not a measure of our certainty that a given void corresponds to a genuine matter underdensity. In the BigMD simulation, this can be checked by direct reference to the density field at the void location. Any void with $\Delta(R_v)<0$ and thus $\nu<0$ certainly qualifies as genuine. However, even those voids which have $\nu>0$ generally still satisfy $\delta_\rmn{min}<0$ and therefore still correspond to genuine underdensities, albeit on some smaller scale. 

Instead the significance $\nu$ represents a measure of the rarity of the density fluctuation associated with a void. Rarer or more extreme voids correspond to more negative values of $\nu$; the most extreme `supervoids' \citep{Szapudi:2015,Finelli:2016} will correspond to $\nu\lesssim-3$. Examples of supervoids drawn from the current catalogue will be presented in \citet{Nadathur:2016b}.

In total we find that once the default selection criterion $\delta_{g,\rmn{min}}<0$ has been applied, only 3\% of all voids in the BigMD mock catalogues fail to satisfy either $\delta_\rmn{min}<0$ or $\Delta(R_v)<0$. This is an acceptable false positive rate, especially since such voids are not easily distinguishable from the larger population on the basis of any directly observable characteristics. Therefore we do not apply any further quality cuts to the DR11 void catalogues.

%==================Fig.: =======================%
\begin{figure*}
\begin{center}
\includegraphics[width=80mm]{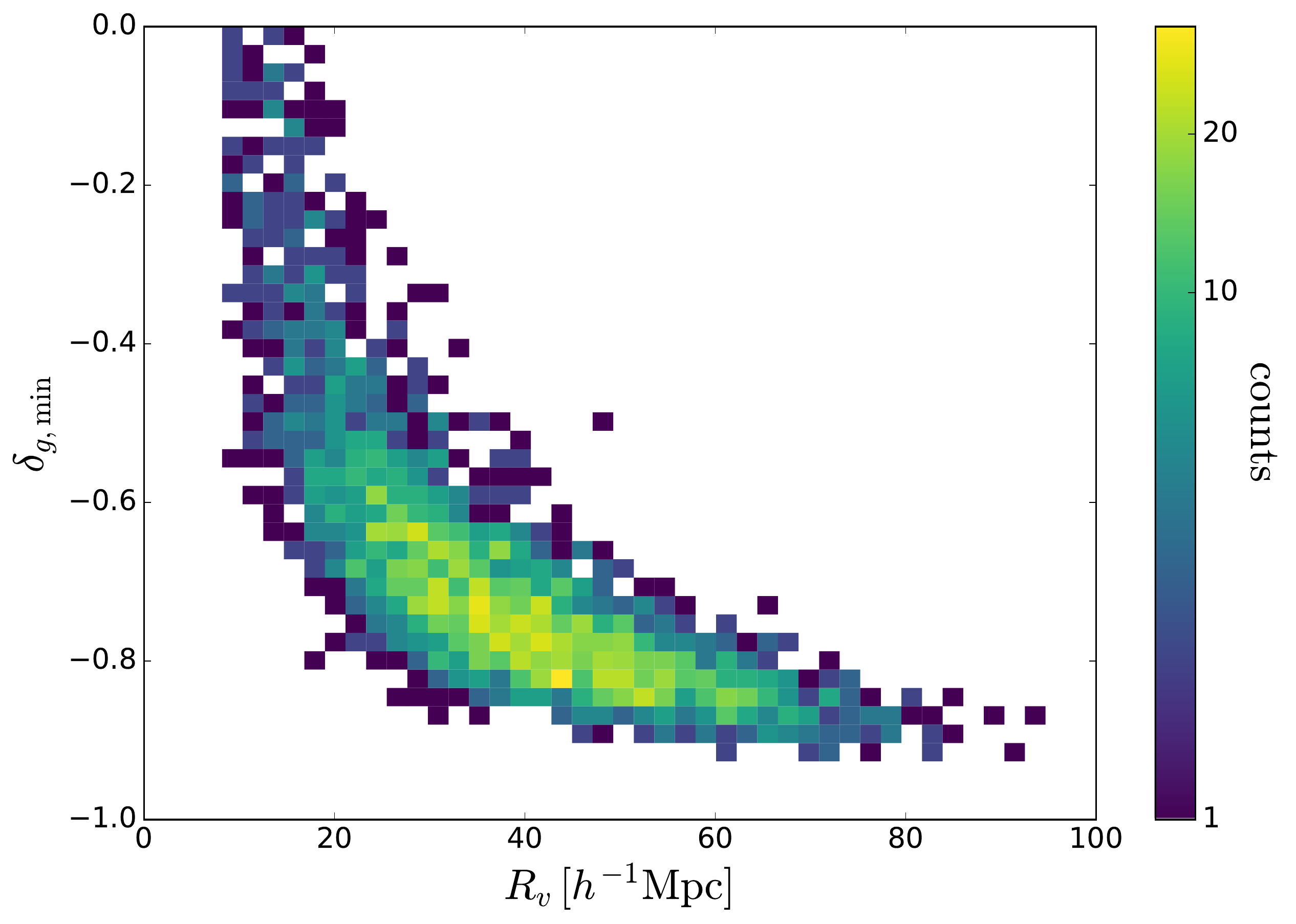}
\includegraphics[width=80mm]{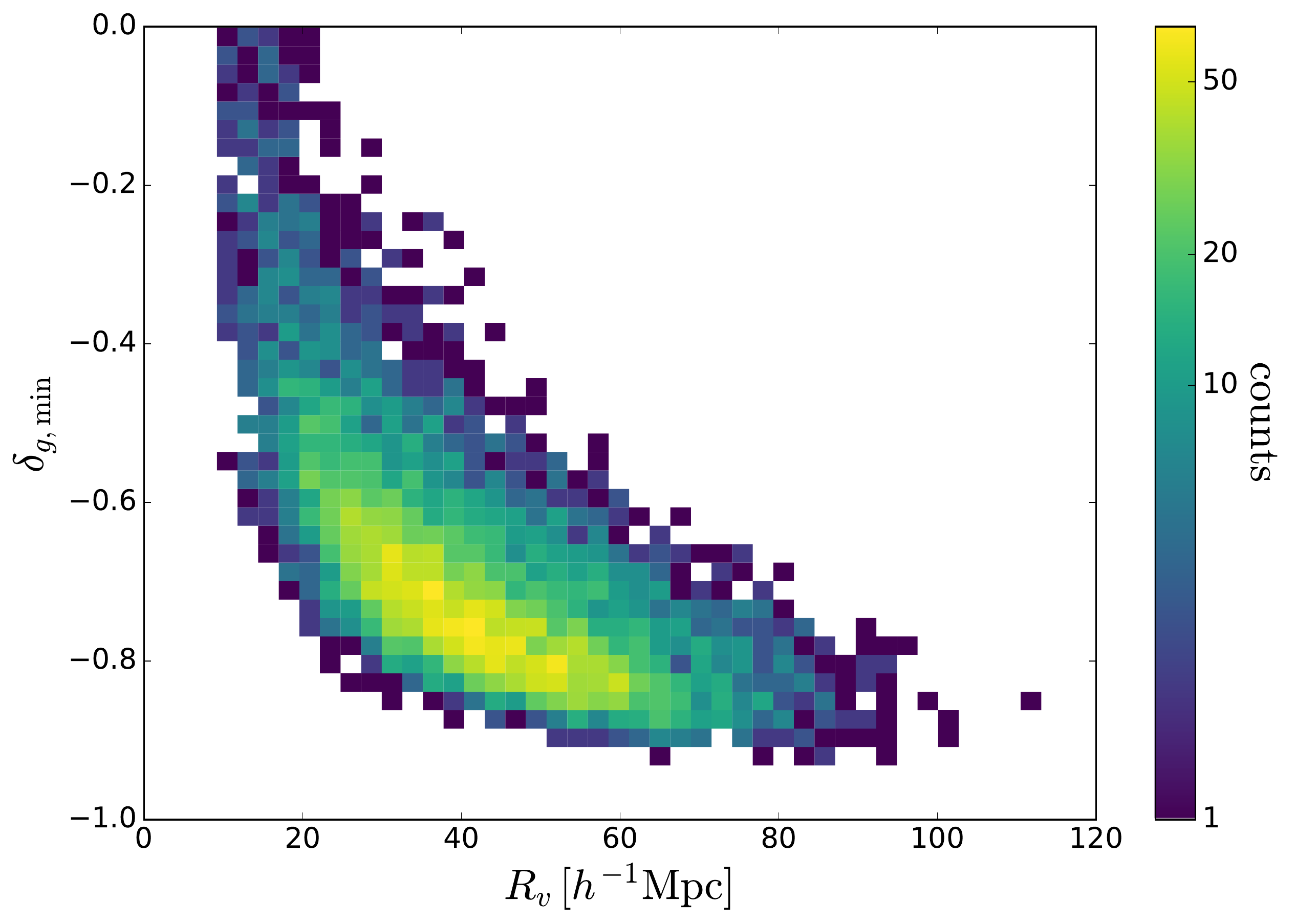}
\caption{The distribution of void effective sizes $R_v$ and minimum galaxy densities $\delta_{g,\mathrm{min}}$ for voids in the LOWZ (left) and CMASS (right) catalogues. For both samples the universal trend \citep{Nadathur:2015b,Nadathur:2015c} towards larger voids containing deeper density minima can be seen. Note the natural density-dependent cutoff at small void sizes imposed by the tessellation resolution.}
\label{fig:deltaRhist}
\end{center}
\end{figure*}
%==================Fig.:=======================%

%==================Fig.: =======================%
\begin{figure*}
\begin{center}
\includegraphics[width=120mm]{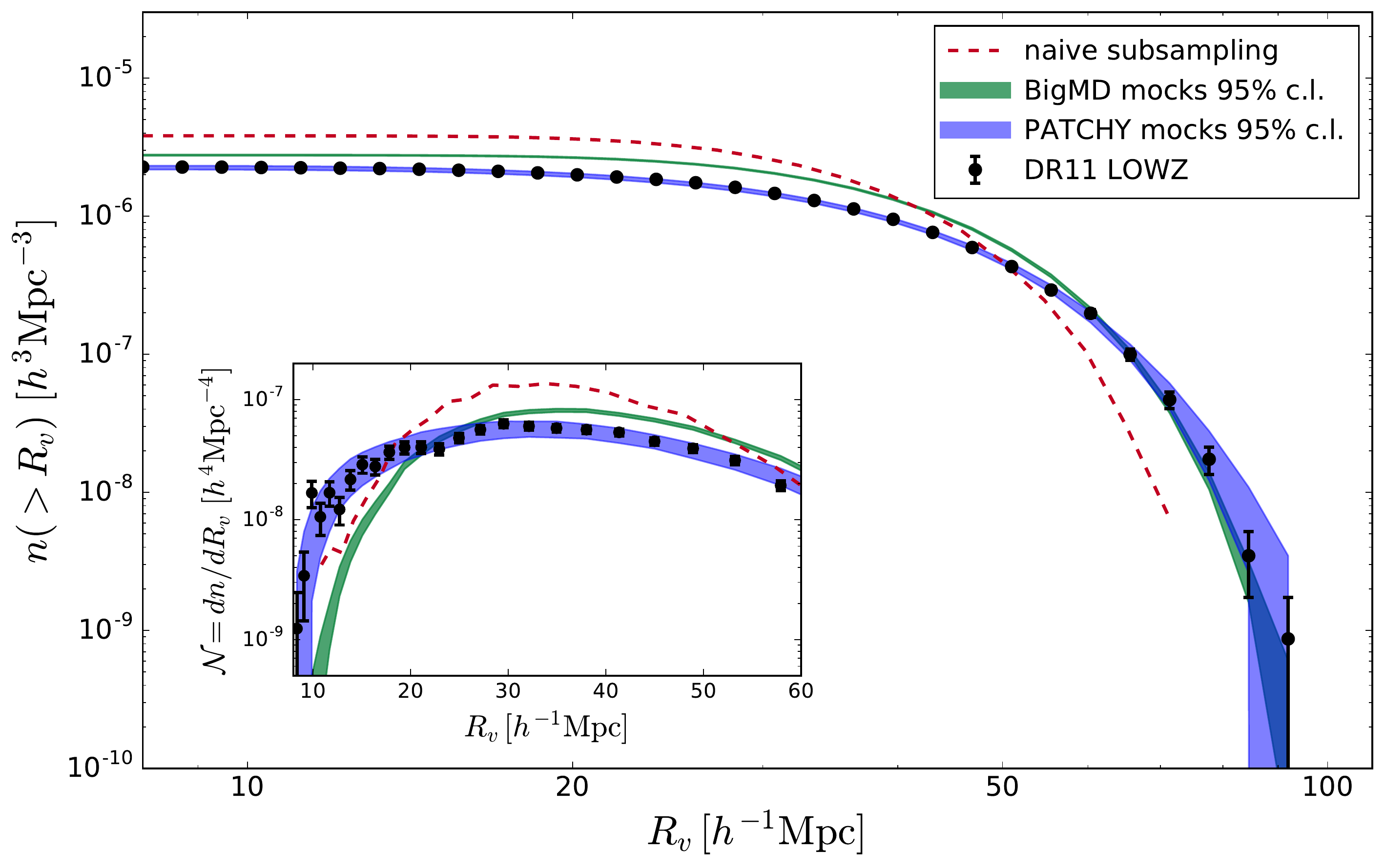}
\caption{The cumulative void abundance as a function of void size, for voids in the LOWZ catalogue (black points). The blue shaded band is the 95\% c.l. range derived from the 1024 {\small PATCHY} mocks matching the galaxy number density, clustering, full survey geometry and selection function. All voids (edge and non-edge) are included. The green shaded band shows the equivalent range for voids in the HOD mocks in the BigMD simulation, which match the galaxy bias and number density but do not account for the survey geometry. The red dashed line is the abundance measured in simulations using naive subsampling of dark matter particles, which matches the galaxy number density but not the bias nor the survey geometry. The black error bars are derived assuming a Poisson distribution in each data bin and are therefore approximations of the true error (blue band). \emph{Inset}: Same as above, but showing the differential number density $\mathcal{N}(R_v)$. The effect of the survey boundary in shifting the void size distribution to smaller sizes is evident.}
\label{fig:LOWZsizes}
\end{center}
\end{figure*}
%==================Fig.:=======================%

%==================Fig.: =======================%
\begin{figure*}
\begin{center}
\includegraphics[width=120mm]{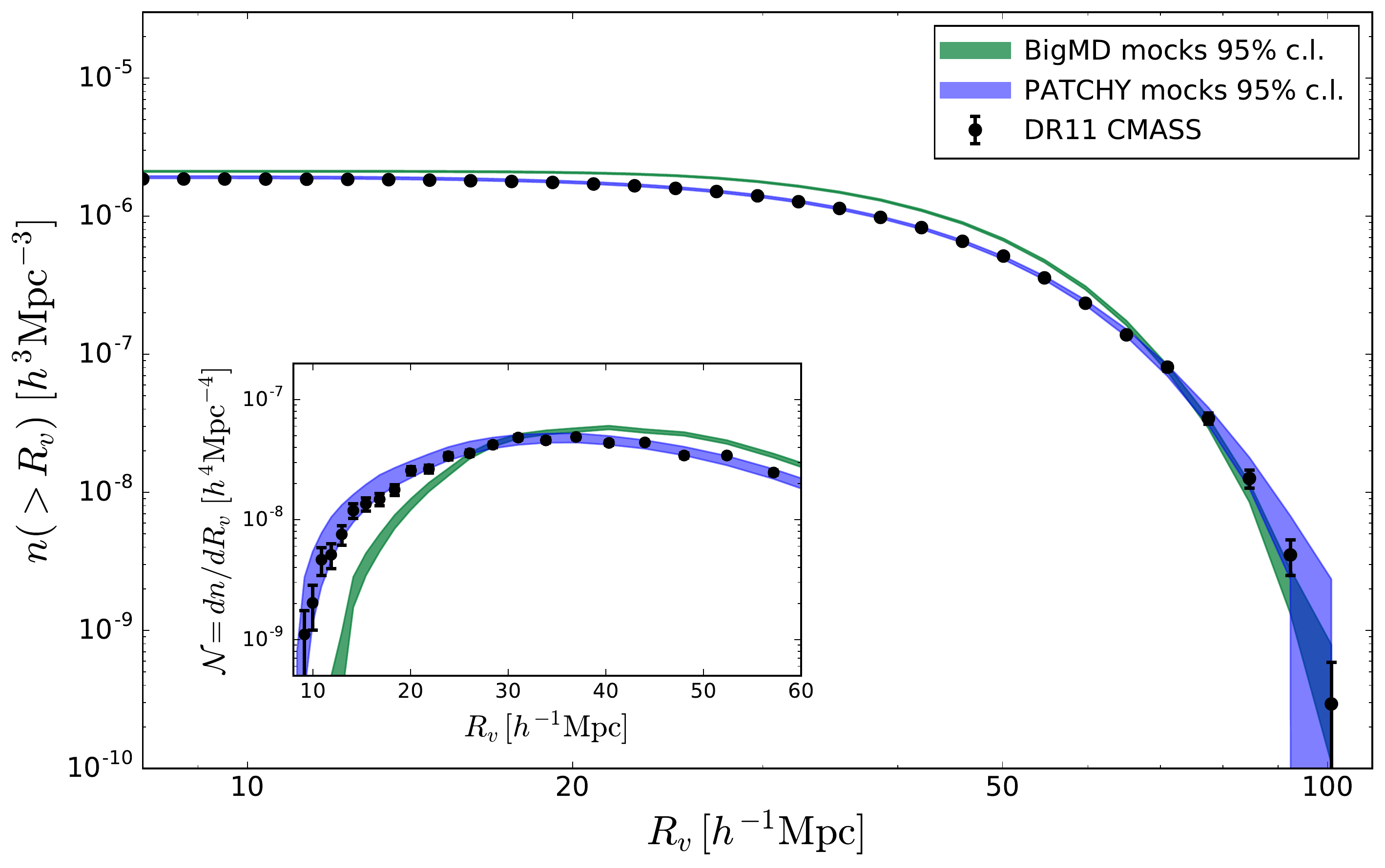}
\caption{Same as in Figure~\ref{fig:LOWZsizes}, but for voids from the CMASS catalogue. Constraints on the void number function are tighter over all scales due to the larger volume of the CMASS survey.}
\label{fig:CMASSsizes}
\end{center}
\end{figure*}
%==================Fig.:=======================%

\section{Results}
\label{sec:results}

We now turn to results from the void catalogues in the DR11 LOWZ and CMASS data. In total we find 8956 independent, non-overlapping voids, of which 6337 are in the CMASS catalogue (North and South combined), and 2619 in LOWZ. This represents the largest public catalogue of galaxy voids to date and thus provides the best statistical constraints on void properties and cosmology, as we discuss below.

\subsection{Size and density distributions}
\label{subsec:sizes_and_densities}

Figure~\ref{fig:deltaRhist} shows the distribution of voids in the $\delta_{g,\rmn{min}}$-$R_v$ plane. For both samples, larger voids contain deeper density minima, consistent with the universal trends observed in simulation \citep{Nadathur:2015b,Nadathur:2015c}. The peak of the distribution occurs at $\delta_{g,\rmn{min}}\sim-0.8$ and $R_v\sim35\,h^{-1}$Mpc for both LOWZ and CMASS.

Averaging over densities, we may describe the void size distribution in terms of the void number function or cumulative comoving number density of voids with effective radius larger than $R_v$, 
\beq
\label{eq:cumnumdens}
n(>R_v) = \int_{R_v}^{\infty}\frac{\rmn{d}n}{\rmn{d}R_v}\rmn{d}R_v.
\eeq
Figures \ref{fig:LOWZsizes} and \ref{fig:CMASSsizes} show the observed void number functions in terms of both $n(>R_v)$ and $\mathcal{N}(R_v)\equiv\rmn{d}n/\rmn{d}R_v$ for the LOWZ and CMASS data respectively. Also shown in these figures are the theoretical predictions for these distributions in the $\Lambda$CDM model, obtained using two different methods. The green shaded band is obtained from the $95\%$ c.l. regions for the void distribution in the BigMD LOWZ and CMASS mocks -- these mocks match the galaxy clustering and bias of the DR11 data, but do not reproduce the survey geometry or selection function $\phi(z)$. The blue shaded band shows the same region for the void distribution in the 1024 {\sm PATCHY} mocks, which include both these effects. The systematic bias introduced by neglecting the survey geometry is clear: while the {\sm PATCHY} mocks provide a very good match to the DR11 voids, the BigMD mocks from a cubic box with periodic boundary conditions are grossly discrepant with both data and {\sm PATCHY} mocks over most of the available void size range.

The reasons for this difference can be understood in terms of survey edge effects. Firstly, the removal of edge galaxies from the tessellation reduces the available survey volume for voids to occupy, as can be seen from Figure~\ref{fig:decslice}. This contributes a small overall offset of the void number density, shifting the survey values downwards. (Such an offset could be removed by renormalising the survey volume to account for the `usable' fraction, but for simplicity and clarity we do not do so here.) Secondly and more importantly, the survey edges truncate the extents of edge voids close to the boundaries, moving them to smaller sizes and thus from right to left in the figure. This changes the shape of the size distribution, as is clear from the inset showing the differential number density $\mathcal{N}$. 

These two boundary effects account for most of the difference between the BigMD voids and those from DR11 and the {\sm PATCHY} mocks at small and intermediate void sizes. We have checked that when regions matching the survey masks for LOWZ and CMASS are cut out of the mocks in the BigMD box and the algorithm reapplied, the resulting distributions match those shown for DR11 and {\sm PATCHY} at these scales. Note also that the BigMD HOD is applied in order to obtain a uniform galaxy number density over the simulation box rather than reproducing the observed selection function in the data. This, together with the increased cosmic variance error at these scales, is likely responsible for the much smaller differences for the very largest voids, where BigMD results slightly underestimate the void number density.

%==================Fig.: =======================%
\begin{figure*}
\begin{center}
\includegraphics[width=80mm]{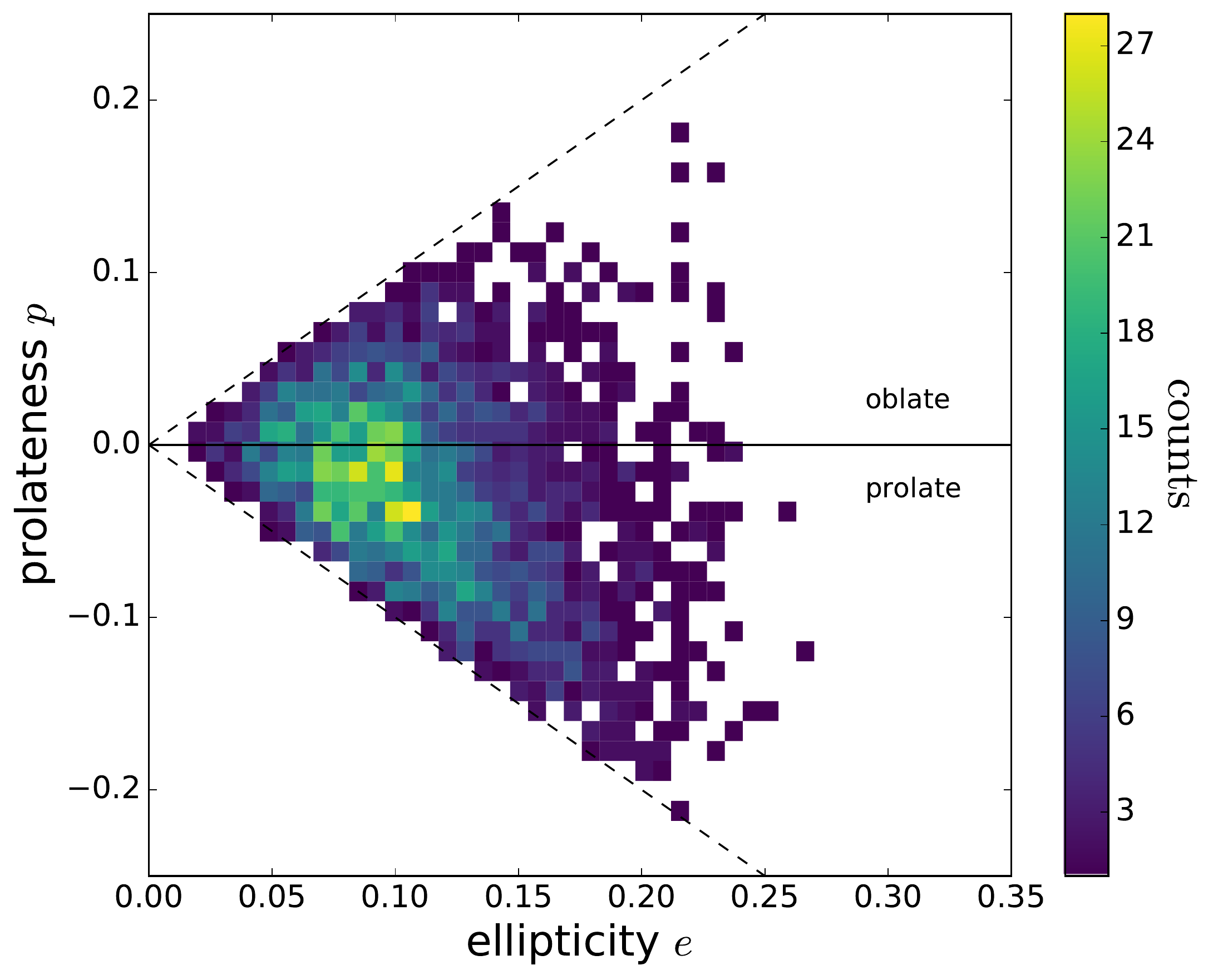}
\includegraphics[width=80mm]{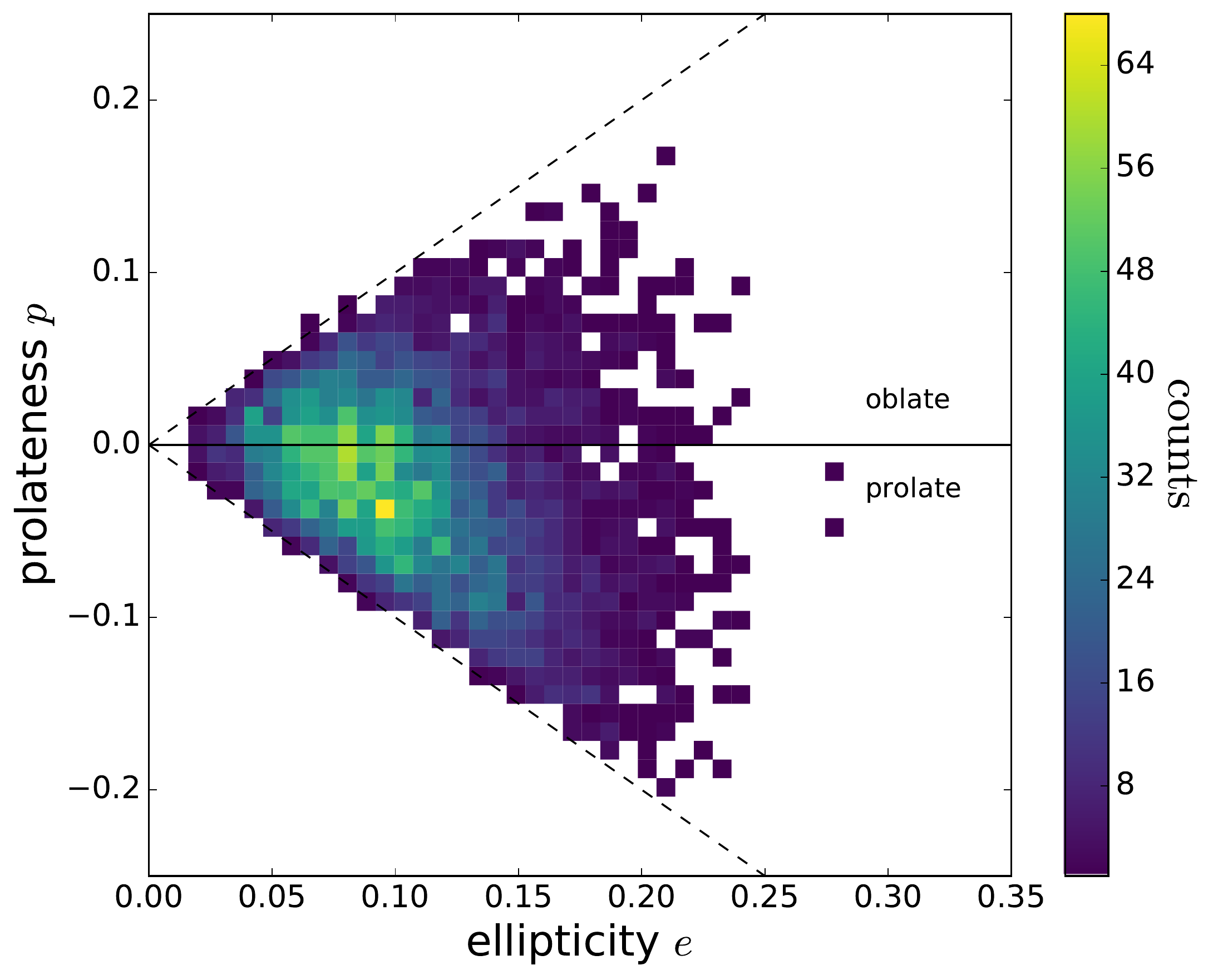}
\caption{The distribution of ellipticity $e$ and prolateness $p$ of voids in the LOWZ (left) and CMASS (right) catalogues. Voids are most appropriately modelled as tri-axial ellipsoids, with an average elongation of the semi-major axis $q=1.47^{+0.20}_{-0.22}$ for LOWZ and $q=1.44^{+0.19}_{-0.20}$ for CMASS (see text). }
\label{fig:e-and-p}
\end{center}
\end{figure*}
%==================Fig.:=======================%

%==================Fig.: =======================%
\begin{figure}
\begin{center}
\includegraphics[width=80mm]{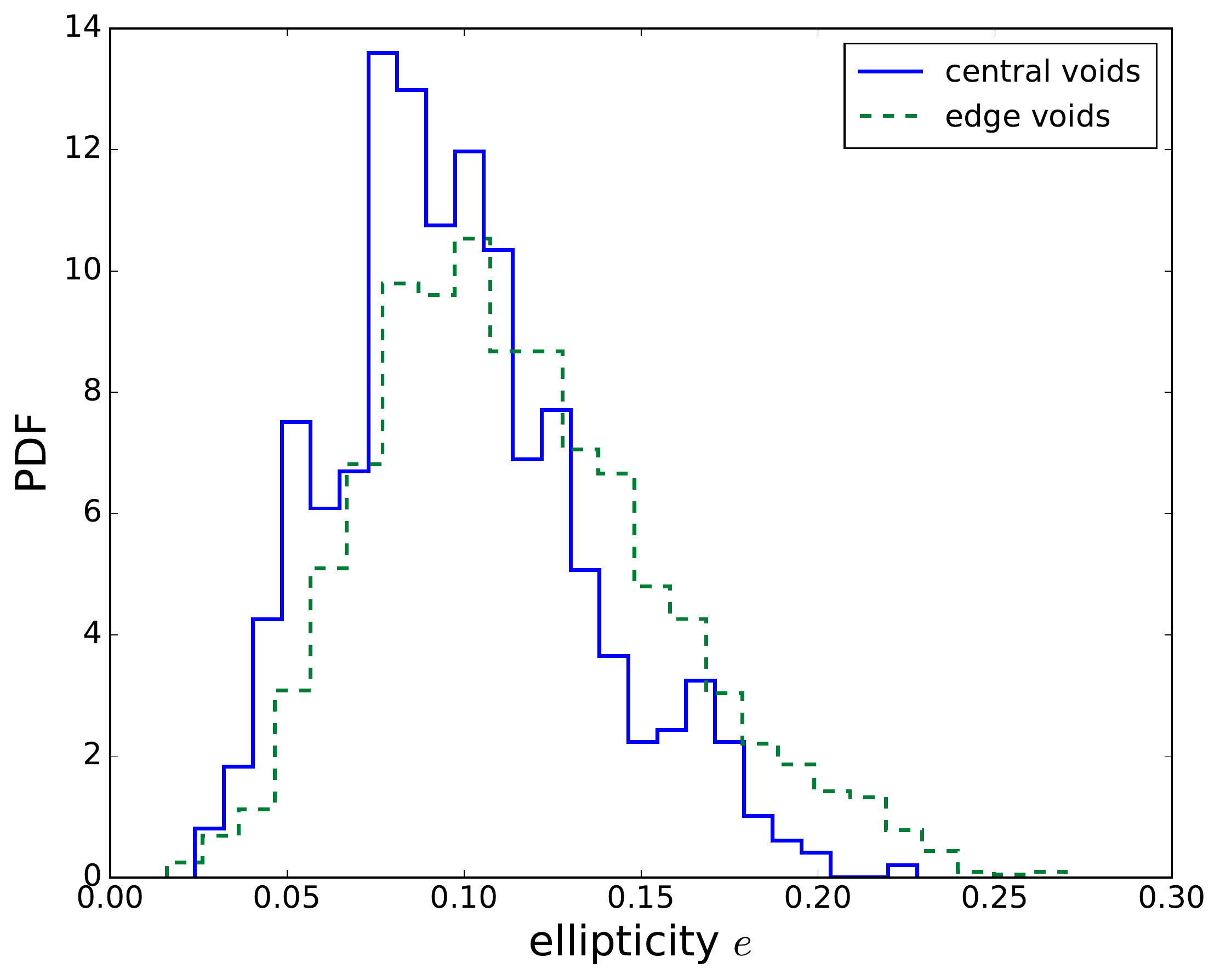}
\caption{The distribution of ellipticity $e$ for central and edge voids in the LOWZ catalogue. Edge voids show a significantly higher ellipticity, due to artificial truncation by the survey edges.}
\label{fig:edge-nonedge-e}
\end{center}
\end{figure}
%==================Fig.:=======================%

Finally, Figure~\ref{fig:LOWZsizes} also shows the void number function for voids found in naively subsampled dark matter particle tracers, where the mean particle number density within the box is set to match that of the LOWZ galaxy sample. This corresponds to the procedure used in a number of studies of the void number function \citep[e.g.,][]{Sutter:2014b,Zivick:2015,Pisani:2015}. As this method ignores the important effects of galaxy bias \citep{Nadathur:2015c,Pollina:2016}, it results in worse estimates of the void size distribution. The relative error from neglecting galaxy bias is greater than that from neglecting the survey geometry.

\subsection{Shapes and ellipticities}
\label{subsec:shapes}

The distribution of DR11 voids in the $e$-$p$ plane is shown in Figure ~\ref{fig:e-and-p}. In general, voids are seen to be tri-axial, with the distribution peaking at ellipticity $e\sim0.1$ and prolateness $p\sim-0.05$, where negative values of $p$ correspond to prolate rather than oblate spheroids. The distribution of values for the ratio $q$ of longest to shortest axis for the equivalent model spheroids is skewed right, with $q=1.47^{+0.20}_{-0.22}$ for voids in LOWZ ($68\%$ c.l.) and $q=1.44^{+0.19}_{-0.20}$ for voids in CMASS. 

This value of the elongation is much smaller than the value of $q=2.6\pm0.4$ found by \citet{Granett:2015} for the voids in the SDSS DR6 Mega-Z photometric galaxy data \citep{Granett:2008a}, which covers a similar sky region and redshift range to the DR11 CMASS North sample and were also identified using {\sm ZOBOV}. There are several possible reasons for this. Firstly, we report individual watershed basins as voids and do not allow merging. As discussed in Sec.~\ref{subsubsec:degeneracies}, voids formed of several watershed regions merged together are generally less symmetric and have higher ellipticity. Secondly, the increased noise due to photo-z redshifts in the Mega-Z data makes it harder to resolve watershed boundaries between voids, leading to increased merging even if none were desired. And thirdly, as discussed by \citet{Flender:2013} and \citet{Granett:2015}, the line-of-sight smearing effect of photo-z uncertainties means that although {\sm ZOBOV} has no directional preference, in practice only those voids that are significantly elongated along the line of sight can be resolved at all, leading to a selection bias in the resulting sample. Note that we find no preferred direction for the elongation, whereas \citet{Granett:2015} report a consistent large elongation along the line of sight.

For DR11 voids, the distribution of the orientation of the long axis with respect to the line of sight direction was found to be isotropic. However, there is a hint from Figure~\ref{fig:decslice} that voids with $\overline\delta_g<0$ might tend to cluster together in space. Taken together with the evidence from Figure~\ref{fig:significance} that such voids also tend to be the most significant, this raises the possibility that several deep matter density fluctuations might line up along a single line of sight. This could increase the stacked ISW \citep{Granett:2008a,Ilic:2013,Hotchkiss:2015a} or lensing \citep{Clampitt:2015} signals along such directions. However, more work is needed to check the feasibility of such a scenario.

Finally, Figure~\ref{fig:edge-nonedge-e} shows the effect of the survey edges on the void ellipticity: the distribution for edge voids is significantly shifted in comparison to that for voids far from the survey boundary. This is clearly because the sizes of such voids are truncated in the direction of the survey edge, resulting in greater asymmetry.

%==================Fig.: =======================%
\begin{figure*}
\begin{center}
\includegraphics[width=85mm]{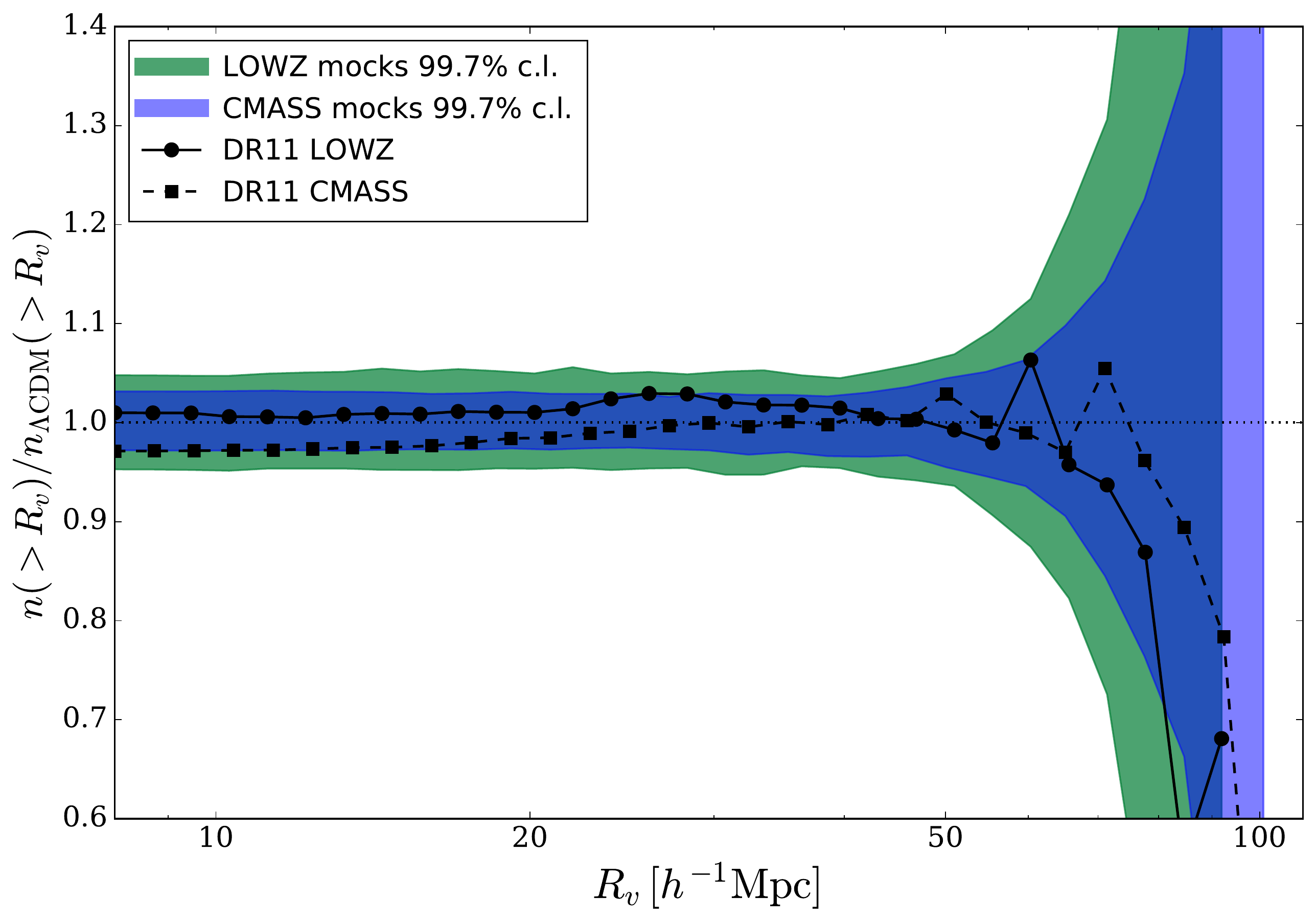}
\includegraphics[width=85mm]{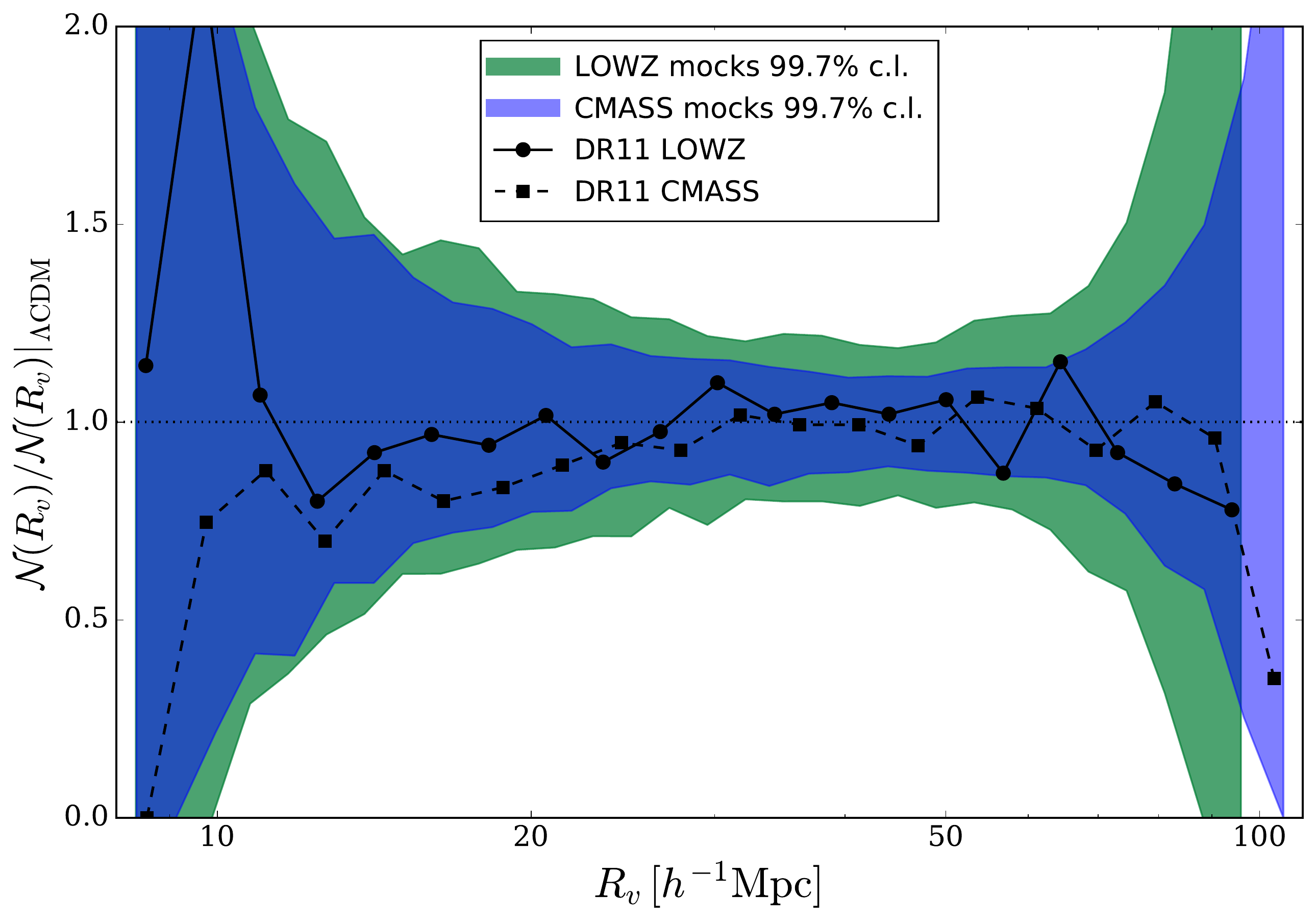}
\caption{Constraints on the deviation of the cumulative (left) and differential (right) void abundance from the $\Lambda$CDM value, as a function of void size. The $\Lambda$CDM mean value and the confidence intervals shown by the shaded contours are determined from the set of 1024 {\small PATCHY} mocks. The agreement between theory and data is excellent at all scales.}
\label{fig:enhancement_nR}
\end{center}
\end{figure*}
%==================Fig.:=======================%

%==================Fig.: =======================%
\begin{figure*}
\begin{center}
\includegraphics[width=85mm]{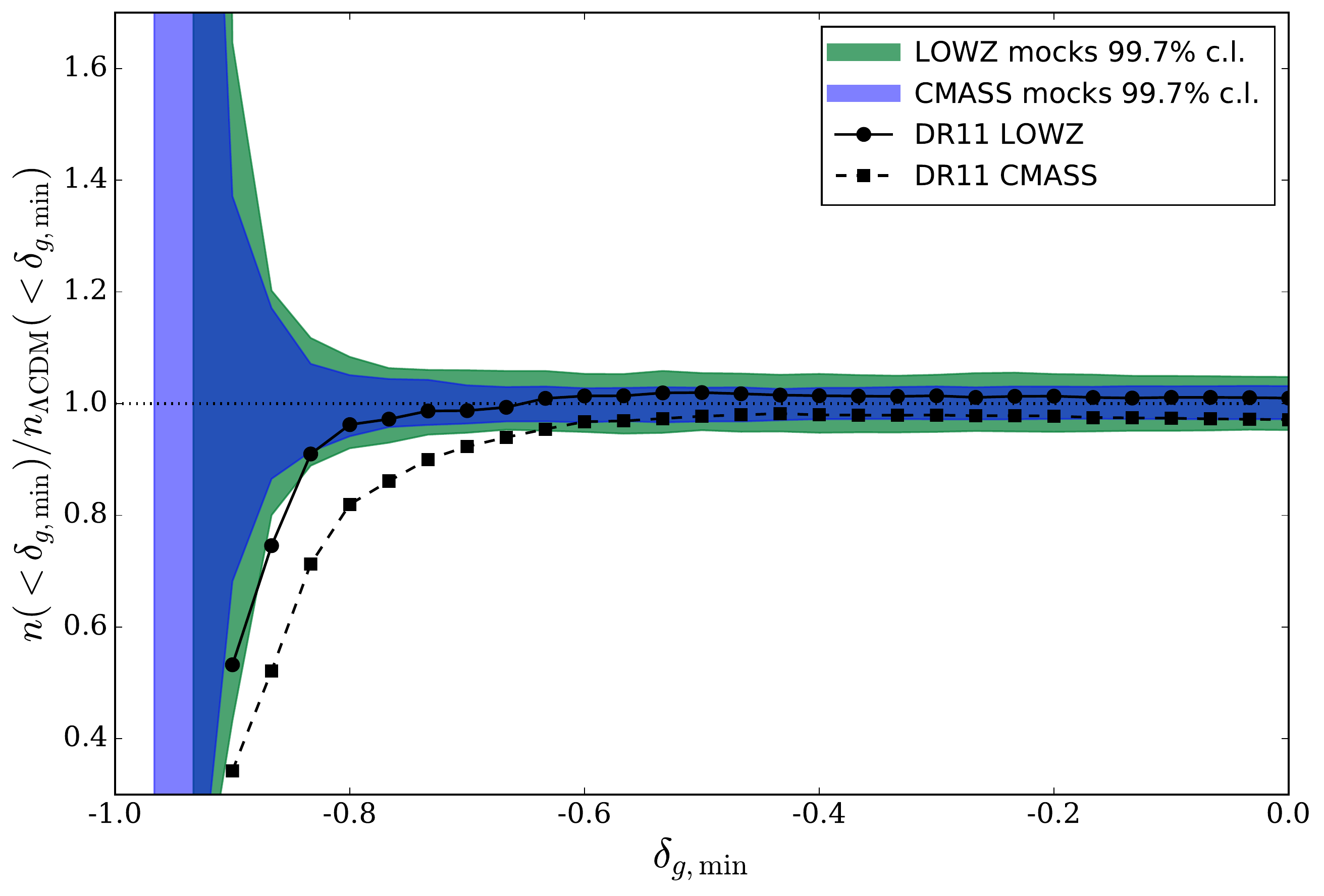}
\includegraphics[width=85mm]{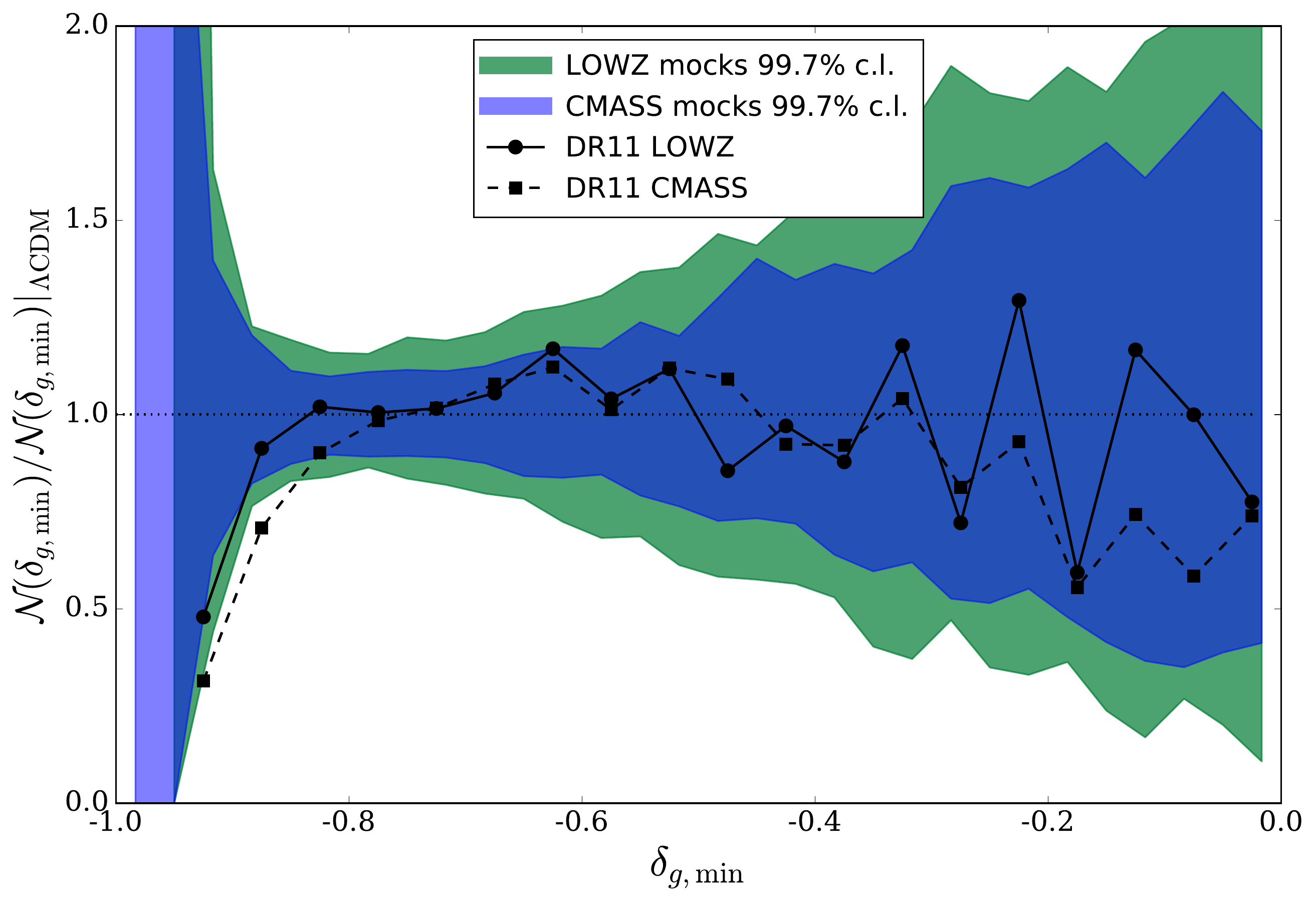}
\caption{As for Figure~\ref{fig:enhancement_nR}, but in this case showing the deviation of the cumulative (left) and differential (right) abundance of voids lying below different minimum galaxy density thresholds $\delta_{g,\rmn{min}}$. Note also that the shaded bands in this case refer to the $99.7\%$ c.l. regions. The CMASS catalogue shows a significant deficit of the deepest voids compared to theoretical predictions.}
\label{fig:enhancement_ndelta}
\end{center}
\end{figure*}
%==================Fig.:=======================%

\section{Testing deviations from $\Lambda$CDM with void statistics}
\label{sec:deviations}

A number of authors have shown that the void size distribution can be a sensitive cosmological probe, capable of differentiating a varying dark energy equation of state \citep{Pisani:2015}, massive neutrino cosmologies \citep{Massara:2015}, modified gravity scenarios \citep{Zivick:2015,Cai:2015}, warm dark matter models \cite{Yang:2014} or coupled dark energy-dark matter models \citep{Pollina:2016} from $\Lambda$CDM. In fact, although this is less commonly studied, the distribution of minimum densities $\delta_{g,\rmn{min}}$ can provide a very similar test. The very large number of voids we find in the DR11 LOWZ and CMASS data, as well as the theoretical mean and error ranges obtained from the set of 1024 {\sm PATCHY} mocks, allow us to constrain the void distributions to unprecedented accuracy and thus place tight constraints on deviations from $\Lambda$CDM.

Figure~\ref{fig:enhancement_nR} shows the constraints on the deviation of the void size distribution from the $\Lambda$CDM value, expressed in terms of both $n(>R_v)$ and $\mathcal{N}(R_v)$. The DR11 data from both LOWZ and CMASS samples are seen to be in excellent agreement with the standard cosmology at all void sizes. Over the void size range $8\lesssim R_v\lesssim60\,h^{-1}$Mpc, deviations of $n(>R_v)$ from the $\Lambda$CDM value are constrained to be $<6\%$ at $95\%$ c.l. for the LOWZ sample at redshifts $0.15<z<0.43$, and to be $<4\%$ for CMASS at redshifts $0.43<z<0.7$. The corresponding constraints on $\mathcal{N}(R_v)$ are naturally less tight, since there is much lower correlation between size bins, however deviations from $\Lambda$CDM are still constrained to be $\lesssim20\%$ for voids between $20\lesssim R_v\lesssim50\,h^{-1}$Mpc for both LOWZ and CMASS.

To put these constraints in perspective, we note that based on simulations of Hu-Sawicki \citep{Hu:2007} $f(R)$ gravity models, \citet{Cai:2015} find that for the parameter value $|f_{R0}|=10^{-6}$, $n(>R_v)$ deviates from its $\Lambda$CDM value by up to $20\%$ at $R_v\sim50\,h^{-1}$Mpc, and that the deviation in $\mathcal{N}(R_v)$ will be correspondingly higher. Similarly, the results of \citet{Pisani:2015} indicate that if the dark energy equation of state is parametrised as $w(z) =w_0 +w_az/(1+z)$ \citep{Chevallier:2001,Linder:2003}, values of $w_a=\pm0.2$ lead to deviations in $n(>R_v)$ of $\sim30\%$ for even smaller voids.

Taken at face value, comparison of such predictions with Figure~\ref{fig:enhancement_nR} indicate that these parameter values can already be excluded on the basis of the data presented here. However, such a direct comparison is complicated by the fact that in making theoretical predictions for the variation of the void size distribution in alternative models, these studies have neither matched the bias of galaxy tracers nor the effects of the survey geometry, which were both shown to be important in Sec.~\ref{subsec:sizes_and_densities}. Further detailed work is therefore required to place quantitative constraints on these scenarios; the level of agreement of the data with $\Lambda$CDM shown here sets the precision standard which must be attained.

Turning to the distribution of void densities, Figure~\ref{fig:enhancement_ndelta} shows the corresponding constraints on the deviations of the distribution functions $n(<\delta_{g,\rmn{min}})$ and $\mathcal{N}(\delta_{g,\mathrm{min}})\equiv\rmn{d}n/\rmn{d}\delta_{g,\rmn{min}}$. Note that in this case contours are shown for the $99.7\%$, or $3\sigma$-equivalent, confidence limits. Here a significant discrepancy with $\Lambda$CDM is seen: the deepest voids, with $\delta_{g,\rmn{min}}\lesssim-0.8$, are much less common in the DR11 data than in the {\sm PATCHY} mocks. This discrepancy is particularly marked for voids in the higher-redshift CMASS sample where the deficit is clearly $>3\sigma$ for several consecutive bins of $\mathcal{N}(\delta_{g,\mathrm{min}})$, but exists to a lesser extent for LOWZ as well.

The most conservative explanation for such a discrepancy is the existence of some undetected systematic effect. However, since the mock catalogues have been constructed to match the two-point and three-point galaxy correlation statistics to high accuracy \citep{Kitaura-DR12mocks:2016}, and exactly the same void-finding pipeline has been applied to both the data and the mock samples, any remaining systematic effect must be subtle. It is possible that this points to some previously unnoticed effect in the {\sm PATCHY} algorithm, which relies on second-order perturbation theory rather than full $N$-body simulations. Alternatively, this may be a consequence of the fact that the void probability function is in principle sensitive to the full hierarchy of correlation functions \citep{White:1979}.

If a subtle systematic effect can be ruled out, an interesting possibility is that such a deficit of the deepest voids is due to some new physics. \citet{Yang:2014} argue that warm dark matter scenarios lead to shallower voids in the matter distribution, which is qualitatively the right effect. A similar effect would be seen for massive neutrino cosmologies \citep{Massara:2015}. However, the effects on the galaxy density within voids may well be smaller. A quantitative study of these scenarios is left to future work.

\section{Conclusions}
\label{sec:conclusions}

We have presented a catalogue of voids in the SDSS-III BOSS DR11 data, from the LOWZ and CMASS galaxy samples. This catalogue contains 8956 independent, non-overlapping voids, making it the largest public catalogue of voids to date. The catalogue contains a wealth of information about each void, including minimum and average galaxy densities, sizes, centre locations, shape parameters and bounding surfaces. Relationships between these observable quantities and the matter content of voids have been calibrated on mock catalogues in a large $N$-body simulation, in order to develop optimal quality control cuts to ensure that as far as possible all voids in the catalogue correspond to genuine underdensities in the matter distribution, with an estimated false positive rate of 3\%. 

In addition to mocks constructed from simulations in cubic boxes with periodic boundary conditions, we have also made use of a suite of 4096 mock galaxy catalogues created by the {\sm PATCHY} algorithm and made available by the SDSS-III collaboration, which match the galaxy clustering for DR11 data as well as the full survey geometry, angular completeness and redshift selection. Comparison of the results in these two cases allows us to isolate the effect of these survey characteristics on void reconstruction, and we have shown that they lead to a strong systematic shift in the distribution of void sizes and ellipticities.

The large number of voids in the catalogue and the use of the {\sm PATCHY} mocks enables tight constraints to be placed on the deviation of the void size distribution from its $\Lambda$CDM expectation value. These constraints are at the level of a few per cent on the ratio $n(>R_v)/n_{\Lambda\rmn{CDM}}(>R_v)$ for void sizes ranging over an order of magnitude. This is significantly smaller than the theoretically predicted deviation of this quantity in a wide range of alternative models, including modified gravity scenarios and models with a varying equation of state. However, the theoretical uncertainties in these predictions are large, partly because they do not account for the systematic bias introduced by the survey geometry that we have shown to be large. Indeed the distorting effect of the survey boundary on void size was discussed by \citet{Sutter:2014c}, but has subsequently been ignored in most theoretical work \citep[however, see][]{Nadathur:2015c,Pollina:2016}. Our current constraints therefore set the precision standard which future theoretical work on voids must attain.

Finally, we have also presented a statistical examination of the distribution of densities within voids, which provides additional cosmological information beyond that present in void sizes alone. Indeed while the void size distribution agrees well with the standard cosmology, we find a deficit of voids with deep density minima in the DR11 data, which exceeds the $3\sigma$ equivalent confidence level limit determined from the {\sm PATCHY} mocks. Since the void-finding pipeline applied to data and mocks is exactly the same, this discrepancy may point to a previously unidentified systematic in the creation of the mock catalogues. Alternatively, it may be a sign of some true physical effect: in this case warm dark matter or massive neutrinos provide possible candidate models, although a thorough quantitative study in future work would be required to test such a hypothesis.

Looking forward, an important application of the void data in this catalogue will be in the measurement of the void weak lensing signal. The data on the void bounding surfaces we have presented here will be particularly important for this, as it enables use of the boundary stacking method introduced by \citet{Cautun:2016}, which has been shown to enhance the shear and magnification signals by a factor of 2. 

Another interesting application will be to stacked void ISW studies, which have so far produced very mixed results. While some authors have reported a high significance detection \citep{Granett:2008a} with an amplitude far in excess of the tiny $\Lambda$CDM expectation \citep{Nadathur:2012,Flender:2013}, others find more marginal or null results \citep{Cai:2013ik,Hotchkiss:2015a,Kovacs:2015,Granett:2015} with amplitudes more consistent with theory. However, until now all such detection attempts have used only $\mathcal{O}(10)$-$\mathcal{O}(100)$ voids. The order-of-magnitude increase in sample size possible with the current catalogue opens the possibility that even the small amplitude signal predicted in $\Lambda$CDM could be detected with high significance.

The void catalogue presented here is publicly available from \url{http://www.icg.port.ac.uk/stable/nadathur/voids/}, and will be updated to include data from DR12 on their public release.

\section{Acknowledgements}

Thanks are due to Shaun Hotchkiss and Robert Crittenden for helpful suggestions and comments on this paper. I acknowledge financial support from an Individual Fellowship of the Marie Sk\l odowska-Curie Actions under the Horizon 2020 Framework of the European Commission, project COSMOVOID.

This work has made use of public data from the SDSS-III collaboration. Funding for SDSS-III has been provided by the Alfred P. Sloan Foundation, the Participating Institutions, the National Science Foundation, and the U.S. Department of Energy Office of Science. The SDSS-III website is \url{http://www.sdss3.org/}. The MultiDark {\sm PATCHY} mocks were produced at the BSC Marenostrum supercomputer, the Hydra cluster at the Instituto de F\'isica Te\'orica UAM/CSIC and NERSC at the Lawrence Berkeley National Laboratory, and were made available via the SDSS-III website.

The BigMultiDark simulations were performed on the SuperMUC supercomputer at the LeibnizRechenzentrum (LRZ) in Munich, using the computing resources awarded to the PRACE project number 2012060963. I am grateful to Gustavo Yepes for preparing some of the density data in this simulation. I also acknowledge use of the {\sm EREBOS}, {\sm THEIA} and {\sm GERAS} clusters at the AIP, for which I thank Stefan Gottl\"ober for granting access and Harry Enke for technical support.
 
%%==================Fig.: =======================%
%\begin{figure*}
%\begin{center}
%\includegraphics[width=85mm]{../../figures/DR11/LOWZ_Isolated_sizes_11x8.pdf}
%\includegraphics[width=85mm]{../../figures/DR11/CMASS_Isolated_sizes_11x8.pdf}
%
%\caption{}
%\label{fig:LOWZsizes}
%\end{center}
%\end{figure*}
%%==================Fig.:=======================%

\bibliographystyle{mn2e}
\bibliography{refs.bib}

\label{lastpage}
\end{document}